\documentstyle[onecolumn,epsfig]{mn}

\newif\ifAMStwofonts

\def \half{\frac{1}{2}}
\def \eps{\epsilon}
\def \veps{\varepsilon}
\def \d{\partial}
\def \lam{\lambda}
\def \non{\nonumber}
\renewcommand{\(}{\left(}
\renewcommand{\)}{\right)}
\newcommand{\df}[2]{ \frac{\partial {#1}}{\partial {#2}} }

\newcommand{\I}{\mbox{\rm i}} 

\ifoldfss
  \ifCUPmtlplainloaded \else
    \NewTextAlphabet{textbfit} {cmbxti10} {}
    \NewTextAlphabet{textbfss} {cmssbx10} {}
    \NewMathAlphabet{mathbfit} {cmbxti10} {} 
    \NewMathAlphabet{mathbfss} {cmssbx10} {} 
  \fi
  \ifAMStwofonts
    \ifCUPmtlplainloaded \else
      \NewSymbolFont{upmath} {eurm10}
      \NewSymbolFont{AMSa} {msam10}
      \NewMathSymbol{\upi}     {0}{upmath}{19}
      \NewMathSymbol{\umu}     {0}{upmath}{16}
      \NewMathSymbol{\upartial}{0}{upmath}{40}
      \NewMathSymbol{\leqslant}{3}{AMSa}{36}
      \NewMathSymbol{\geqslant}{3}{AMSa}{3E}

       \let\le=\leqslant
       \let\ge=\geqslant
    \fi
  \fi
\fi 

\ifnfssone
  \newmathalphabet{\mathit}
  \addtoversion{normal}{\mathit}{cmr}{m}{it}
  \addtoversion{bold}{\mathit}{cmr}{bx}{it}
  \newmathalphabet{\mathbfit} 
  \addtoversion{normal}{\mathbfit}{cmr}{bx}{it}
  \addtoversion{bold}{\mathbfit}{cmr}{bx}{it}
  \newmathalphabet{\mathbfss} 
  \addtoversion{normal}{\mathbfss}{cmss}{bx}{n}
  \addtoversion{bold}{\mathbfss}{cmss}{bx}{n}
  \ifAMStwofonts
    \ifCUPmtlplainloaded \else
      \UseAMStwoboldmath
      \makeatletter
      \new@mathgroup\upmath@group
      \define@mathgroup\mv@normal\upmath@group{eur}{m}{n}
      \define@mathgroup\mv@bold\upmath@group{eur}{b}{n}
      \edef\UPM{\hexnumber\upmath@group}
      \new@mathgroup\amsa@group
      \define@mathgroup\mv@normal\amsa@group{msa}{m}{n}
      \define@mathgroup\mv@bold\amsa@group{msa}{m}{n}
      \edef\AMSa{\hexnumber\amsa@group}
      \makeatother
      \mathchardef\upi="0\UPM19
      \mathchardef\umu="0\UPM16
      \mathchardef\upartial="0\UPM40
      \mathchardef\leqslant="3\AMSa36
      \mathchardef\geqslant="3\AMSa3E

       \let\le=\leqslant
       \let\ge=\geqslant
    \fi
  \fi
\fi 

\ifnfsstwo
  \DeclareMathAlphabet{\mathbfit}{OT1}{cmr}{bx}{it}
  \SetMathAlphabet\mathbfit{bold}{OT1}{cmr}{bx}{it}
  \DeclareMathAlphabet{\mathbfss}{OT1}{cmss}{bx}{n}
  \SetMathAlphabet\mathbfss{bold}{OT1}{cmss}{bx}{n}
  \ifAMStwofonts
    \ifCUPmtlplainloaded \else
      \DeclareSymbolFont{UPM}{U}{eur}{m}{n}
      \SetSymbolFont{UPM}{bold}{U}{eur}{b}{n}
      \DeclareSymbolFont{AMSa}{U}{msa}{m}{n}
      \DeclareMathSymbol{\upi}{0}{UPM}{"19}
      \DeclareMathSymbol{\umu}{0}{UPM}{"16}
      \DeclareMathSymbol{\upartial}{0}{UPM}{"40}
      \DeclareMathSymbol{\leqslant}{3}{AMSa}{"36}
      \DeclareMathSymbol{\geqslant}{3}{AMSa}{"3E}

       \let\le=\leqslant
       \let\ge=\geqslant
    \fi
  \fi
\fi 

\ifCUPmtlplainloaded \else
  \ifAMStwofonts \else 
    \def\upi{\pi}
    \def\umu{\mu}
    \def\upartial{\partial}
  \fi
\fi

\title[On the $r$-mode spectrum of relativistic stars in the
low-frequency approximation]{On the $r$-mode spectrum of relativistic
  stars in the low-frequency approximation}

\author[J.~Ruoff and K.D.~Kokkotas]
{Johannes Ruoff and Kostas D.~Kokkotas \\
  Department of Physics, Aristotle University of Thessaloniki,
  Thessaloniki 54006, Greece}

\date{Accepted ???? Month ??.
      Received 2001 Month ??;
      in original form 2001 Month ??}
\pagerange{\pageref{firstpage}--\pageref{lastpage}}
\pubyear{2001}

\begin{document}

\maketitle

\label{firstpage}

\begin{abstract}
  The axial modes for non-barotropic relativistic rotating neutron
  stars with uniform angular velocity are studied, using the
  slow-rotation formalism together with the low-frequency
  approximation, first investigated by Kojima. The time independent
  form of the equations leads to a singular eigenvalue problem, which
  admits a continuous spectrum. We show that for $l=2$, it is
  nevertheless also possible to find discrete mode solutions (the
  $r$-modes).  However, under certain conditions related to the
  equation of state and the compactness of the stellar model, the
  eigenfrequency lies inside the continuous band and the associated
  velocity perturbation is divergent; hence these solutions have to be
  discarded as being unphysical. We corroborate our results by
  explicitly integrating the time dependent equations. For stellar
  models admitting a physical $r$-mode solution, it can indeed be
  excited by arbitrary initial data. For models admitting only an
  unphysical mode solution, the evolutions do not show any tendency to
  oscillate with the respective frequency. For higher values of $l$ it
  seems that in certain cases there are no mode solutions at all.
\end{abstract}


\section{Introduction}

Immediately after the discovery of the $r$-modes being generically
unstable with respect to gravitational-wave emission (Andersson 1998a;
Friedman \& Morsink 1998), it was suggested that they may cause the
newly born neutron stars to spin down via the emission of
gravitational waves (Lindblom, Owen \& Morsink 1998; Andersson,
Kokkotas \& Schutz 1999). Because of their surprisingly fast growth
times, $r$-modes should be able to slow down a hot and rapidly
spinning newly born neutron star during the first months of its
existence. There is also work suggesting that the $r$-mode
instability might be relevant for old neutron stars in binary systems.
This potential relevance for astrophysics has attracted the interest
of both the relativity and the astrophysical community on various
aspects of this subject. For an exhaustive upto-date review, see for
instance Andersson \& Kokkotas (2001) and Friedman \& Lockitch (2001).

Most of the recent work on $r$-modes is based on Newtonian
calculations under the assumption of slow rotation, and the effects of
gravitational radiation are incorporated through the quadrupole
formula. However, it is clear that for a complete and quantitatively
correct understanding, one has to use the framework of general
relativity. Still the slow-rotation approximation is well justified,
since the angular velocity of even the fastest spinning known pulsar
corresponds to a rotational expansion parameter of $\veps =
\Omega/\sqrt{M/R^3}\approx 0.3$. The full set of equations in the
slow-rotation limit was first given by Chandrasekhar \& Ferrari (1991)
for the axisymmetric case, and by Kojima (1992) for the general case.

In the non-rotating case, the perturbation equations are decoupled
with respect to the harmonic index $l$ and degenerate with respect to
the azimuthal index $m$. Furthermore the oscillation modes can be
split into two independent sets, which are characterized by their
behaviour under parity transformation.  The {\em polar} (or {\em
  spheroidal}) modes transform according to $(-1)^l$, whereas the {\em
  axial} (or {\em toroidal}) modes according to $(-1)^{l+1}$. For a
non-rotating perfect fluid star, the only possible fluid oscillations
are the spheroidal $f$- and $p$-modes. For non-barotropic stars,
i.e.~stars with a temperature gradient or a composition gradient,
there exists another family of modes, the $g$-modes, where the main
restoring force is gravity. All axial perturbations of non-rotating
perfect fluid stars have zero frequency, i.e.~they represent
stationary currents. In the non-barotropic case, this zero-frequency
space consists only of the axial $r$-modes, while for barotropic stars
\footnote{Following Lockitch et al. (2001) we call a stellar model
  {\em barotropic} if the unperturbed configuration obeys the same
  one-parameter equation of state as the perturbed configuration.}, it
also includes the polar $g$-modes, since they require a temperature
gradient for their existence.

As the star is set into rotation, the picture changes. In the slow
rotation approximation, the $m$-degeneracy is removed and the polar
modes with index $l$ are coupled to the axial modes with indices $l\pm
1$ and vice versa. Furthermore, the rotation imparts a finite
frequency to the zero-frequency perturbations of the non-rotating
stars. In non-barotropic stars, those modes, whose restoring force is
the Coriolis force, all have axial parity. However, as has been first
pointed out in the Newtonian framework by Lockitch \& Friedman (1999),
for barotropic stars the rotationally restored (inertial) modes are
generically hybrids, whose limit in the non-rotating case are
mixtures of axial and polar perturbations.

If one focuses only on the $r$-mode, whose frequency is proportional
to the star's angular velocity $\Omega$, one can order the
perturbation variables in powers of $\Omega$. Kojima (1997, 1998) used
this low-frequency approximation, sometimes also called slow-motion
approximation (Schumaker \& Thorne, 1983), to show that the purely
axial modes of (non-barotropic) stellar models can be described by a
single second-order ODE.  This eigenvalue problem, however, proves to
be singular, since it is possible for the highest derivate to vanish
at some value of the radial coordinate inside the star. Kojima
(1997,1998) then argued that this singular structure should give rise
to a continuous spectrum. This has been put on a rigorous mathematical
footing by Beyer \& Kokkotas (1999). The appearance of a continuous
spectrum can be explained as follows. It is well known that in the
Newtonian limit, the eigenfrequency of the $r$-mode for an inertial
observer is given by
\begin{equation}\label{freqNewt}
  \sigma_N = -m\Omega\(1  - \frac{2}{l(l+1)}\)\;.
\end{equation}
A first relativistic correction can be obtained by using the
relativistic Cowling approximation, which consists in neglecting all
metric perturbations. In this case, the only correction comes from the
frame dragging $\omega$, which is a function of the radial coordinate
$r$, thus leading to an $r$-dependent oscillation frequency of each
fluid layer:
\begin{equation}\label{freqCowl}
  \sigma_C(r) = -m\Omega\left[1 - \frac{2}{l(l+1)}
  \(1 - \frac{\omega(r)}{\Omega}\)\right]\;.
\end{equation}
Instead of a single frequency, there is now a continuous band of
frequencies, whose boundaries are determined by the values of the
frame dragging $\omega(r)$ at the centre and the surface of the star.

However, it has been pointed out (Beyer \& Kokkotas 1999) that the
existence of the continuous spectrum might be just an artefact of the
too restricted low-frequency approximation. With the inclusion of
gravitational radiation effects, the frequencies become
complex-valued, thus potentially removing the singular structure of
Kojima's equation. But even in the case of real-valued frequencies, it
has been recently shown by Lockitch, Andersson \& Friedman (2001) that
for a non-barotropic uniform density model, in addition to the
continuous spectrum there also exists a single mode solution with the
eigenfrequency lying outside the continuous band. It is this mode that
represents the relativistic $r$-mode for non-barotropic stars.

In this paper, we extend the search of $r$-mode solutions to stars
with various polytropic and realistic equations of state (EOS). We
shall show that in addition to the continuous part, the eigenvalue
equation always admits a single mode solution, at least for $l=2$.
However, for some stellar models, depending on the polytropic index
$n$ and on the compactness, the frequency of this solution lies inside
the continuous band and is associated with a divergence in the fluid
perturbation at the singular point. This is clearly not acceptable,
and therefore we have to discard such solutions as being unphysical.
As a logical consequence, we conclude that in those cases, there do
not exist any $r$-modes, at least within the low-frequency
approximation. In an independent work, Yoshida (2001) has come to a
similar conclusion. He showed that even when studied in the
post-Newtonian approximation, some polytropic models do not admit any
$r$-modes.

For realistic EOS, the existence of $r$-modes depends on the average
polytropic index of the high-density regime. For very stiff EOS, the
neutron star models can exhibit $r$-modes throughout the complete
physically acceptable mass range, whereas for the very soft EOS, none
of the neutron star models does. For EOS in the intermediate range,
the existence of $r$-modes depends on the compactness of the stellar
model. In addition to the mode calculations, we also use the time
depend form of the equations. For those cases where we can find a
physical $r$-mode solution, the Fourier spectrum of the time evolution
does indeed show a peak at the appropriate frequency, whereas for
those cases where we only have the unphysical mode, it does not.

\section{Mathematical formulation}

Assuming that the star is slowly rotating with a uniform angular
velocity $\Omega$, we neglect all terms of order higher than $\Omega$.
In this approximation, the star remains spherical, because the
deformation due to centrifugal forces is of order $\Omega^2$. Thus,
the metric can be written in the form
\begin{equation}
  ds^2_0 = -e^{2\nu} dt^2 + e^{2\lam} dr^2
  + r^2\(d\theta^2 + \sin^2\theta d\phi^2\)
  - 2\omega r^2\sin^2\theta dtd\phi\;,
\end{equation}
where $\nu$, $\lam$ and the ``frame dragging'' $\omega$ are
functions of the radial coordinate $r$ only. With the neutron star matter
described by a perfect fluid with pressure $p$, energy density
$\eps$, and four-velocity
\begin{equation}
  \(u^t, u^r, u^\theta, u^\phi\) = \(e^{-\nu}, 0, 0, \Omega e^{-\nu}\)\;,
\end{equation}
the Einstein equations, together with a one-parameter equation of
state $p=p(\eps)$, yield the well-known TOV equations plus an extra
equation for the function $\varpi$, defined as
\begin{equation}
  \varpi := \Omega - \omega\;.
\end{equation}
To linear order, this equation is
\begin{equation}\label{drag}
  \varpi'' - \(4\pi re^{2\lam}(p + \eps) - \frac{4}{r}\)\varpi'
  - 16\pi e^{2\lam}\(p + \eps\)\varpi = 0\;.
\end{equation}
In the exterior, it reduces to
\begin{equation}
  \varpi'' + \frac{4}{r}\varpi' = 0\;,
\end{equation}
for which we have the solution (Hartle 1967)
\begin{equation}\label{oex}
  \varpi = \Omega - \frac{2J}{r^3}\;,
\end{equation}
with $J$ being the total angular momentum of the neutron star.
Equation (\ref{drag}) has to be integrated from the centre of the star
to its surface $R$, where it has to match smoothly the exterior
solution (\ref{oex}). With the angular momentum given by (Hartle 1967,
Glendenning 1997)
\begin{equation}
  J = \frac{8\pi}{3}\int_0^R{r^4e^{\lam-\nu}\(p + \eps\)\varpi}\,dr\;,
\end{equation}
the matching condition becomes
\begin{equation}
  R^4\varpi'(R) = 6J\;,
\end{equation}
and with Eq.~(\ref{oex})
\begin{equation}\label{mcond}
  \varpi'(R) = \frac{3}{R}(\Omega - \varpi(R))\;.
\end{equation}

If we focus on pure axial perturbations, the perturbed metric can be
written in the following form:
\begin{equation}\label{metric}
  ds^2 = ds^2_0 + 2\sum_{l,m}\(h_0^{lm}(t,r)dt + h_1^{lm}(t,r)dr\)
  \(-\frac{\d_\phi Y_{lm}}{\sin\theta}d\theta 
  + \sin\theta\d_\theta Y_{lm}d\phi\)\;,
\end{equation}
where $Y_{lm} = Y_{lm}(\theta,\phi)$ denote the scalar spherical
harmonics.  In addition, the axial component of the fluid velocity
perturbation can be expanded as
\begin{equation}\label{fluid}
  4\pi(p + \eps)\(\delta u^\theta, \delta u^\phi\)
  = e^{\nu}\sum_{l,m}U^{lm}(t,r)
  \(-\frac{\d_\phi Y_{lm}}{\sin\theta}, \sin\theta\d_\theta Y_{lm}\)\;.
\end{equation}
Einstein's field equations then reduce to four equations for the three
variables $h_0^{lm}$, $h_1^{lm}$ and $U^{lm}$ (Kojima 1992).

As an alternative, we can use the ADM-formalism (Arnowitt, Deser \&
Misner 1962) to derive the evolution equations for the axial
perturbations. The usefulness of this formalism for the perturbation
equations of non-rotating neutron stars has been showed in Ruoff
(2001). This formalism can be taken over to rotating stars and we can
deduce equations describing the evolution of purely axial oscillations
of slowly rotating neutron stars in terms of metric and extrinsic
curvature variables. We should mention that our derivation starts with
the complete set of perturbations, including both polar and axial
perturbations. Only at the end do we neglect the coupling between the
two parities and focus only on the axial equations. In the
Regge-Wheeler gauge, there are just 2 non-vanishing axial metric
perturbations and 2 axial extrinsic curvature components. In the
notation of Ruoff (2001), they metric components are given by
\begin{eqnarray}
  \(\beta_\theta,\,\beta_\phi\) &=& e^{\nu-\lam}\sum_{l,m}K_6^{lm}
  \(-\frac{\d_\phi Y_{lm}}{\sin\theta}, \sin\theta\d_\theta Y_{lm}\)\;,\\
  \(h_{r\theta},\,h_{r\phi}\) &=& e^{\lam-\nu}\sum_{l,m}V_4^{lm}
  \(-\frac{\d_\phi Y_{lm}}{\sin\theta}, \sin\theta\d_\theta Y_{lm}\)\;,
\end{eqnarray}
and the extrinsic curvature components read
\begin{eqnarray}
  \(k_{r\theta},\,k_{r\phi}\) &=& \half e^{\lam}\sum_{l,m}K_3^{lm}
  \(-\frac{\d_\phi Y_{lm}}{\sin\theta}, \sin\theta\d_\theta Y_{lm}\)\;,\\
  \(\begin{array}{cc}
    k_{\theta\theta} & k_{\theta\phi}\\
    k_{\phi\theta} & k_{\phi\phi}\end{array}\) &=&
  \half e^{-\lam}\sum_{l,m}K_6^{lm}\sin\theta
  \(\begin{array}{cc}
    -\sin\theta^{-2}X_{lm} & W_{lm}\\
    W_{lm} & X_{lm}\end{array}\)\;,
\end{eqnarray}
where $W_{lm}$ and $X_{lm}$ are abbreviations for
\begin{eqnarray}
  W_{lm} &=& \(\d^2_\theta + l(l+1)\)Y_{lm}\;,\\
  X_{lm} &=& 2\(\d_\theta - \cot\theta\)\d_\phi Y_{lm}\;.
\end{eqnarray}
For the fluid velocity perturbation, we use the covariant form
\begin{equation}
  \(\delta u_\theta, \delta u_\phi\) = e^{-\lam}\sum_{l,m}u_3^{lm}(t,r)
  \(-\frac{\d_\phi Y_{lm}}{\sin\theta}, \sin\theta\d_\theta Y_{lm}\)\;.
\end{equation}
The relation to the expansions (\ref{metric}) and (\ref{fluid}) is
given by (from now on we omit the indices $l$ and $m$):
\begin{eqnarray}
  h_0 &=& e^{\nu-\lam} K_6\;,\\
  h_1 &=& e^{\lam-\nu} V_4\;,\\
  U &=& 4\pi e^{-\lam-\nu}(p + \eps)\(u_3 - K_6\)\;.
\end{eqnarray}
We obtain the following quite simple set of evolution equations
for the variables $V_4$, $K_3$, $K_6$ and $u_3$:
\begin{eqnarray}
  \label{V4}
  \(\df{}{t} + \I m\omega\)V_4 &=& e^{2\nu-2\lam}
  \left[K_6' + \(\nu' - \lam' - \frac{2}{r}\)K_6 - e^{2\lam}K_3\right]\;,\\
  \label{K3}
  \(\df{}{t} + \I m\omega\)K_3 &=& \frac{l(l+1) -2}{r^2}V_4
  + \frac{2\I m}{l(l+1)}\omega'e^{-2\lam}K_6\;,\\
  \label{K6}
  \(\df{}{t} + \I m\omega\)K_6 &=& V_4' - \frac{\I mr^2}{l(l+1)}
  \left[\omega'K_3 - 16\pi(\Omega - \omega)(p + \eps)u_3\right]\;,\\
  \label{u3}
  \(\df{}{t} + \I m\Omega\)u_3 &=& \frac{2\I m(\Omega-\omega)}{l(l+1)}
  \(u_3 - K_6\)\;.
\end{eqnarray}
Furthermore, we have one momentum constraint:
\begin{equation}\label{MC_odd}
  16\pi(p + \eps)u_3 = K_3' + \frac{2}{r}K_3
  - \frac{l(l+1) - 2}{r^2}K_6
  - \frac{2\I m\omega'}{l(l+1)}e^{-2\nu}V_4\;.
\end{equation}
These equations are completely equivalent to the axial parts of
Eqs.~(20), (24), (25) and (27) of Kojima (1992) when the coupling to
the polar equations is neglected.

\section{Low-frequency approximation}

The above evolution equations should not only describe the $r$-modes
but also the axial $w$-modes, which have much higher oscillation
frequencies.  If we want to focus on the $r$-modes only, we can use
the fact that from Eq.~(\ref{freqNewt}), it follows that the
$r$-mode frequency $\sigma$ is proportional to the star's angular
velocity $\Omega$. Hence, we can require that in our evolution
equations (\ref{V4}) -- (\ref{u3}), the time derivative $\d_t$ be
proportional to the $r$-mode frequency $\sigma$ or, equivalently, to
$\Omega$ (Kojima 1997, 1998).  In this case, we can order the
perturbation variables in powers of $\Omega$ (Lockitch et al.~2001) as
\begin{eqnarray}
  \label{eq:order}
  u_3, K_3, K_6  &\sim& O(1)\;,\non\\
  V_4 &\sim& O(\Omega)\;.
\end{eqnarray}
Keeping only the lowest order terms, we can neglect terms proportional
to $V_4$ in the evolution equation (\ref{V4}) and in the constraint
(\ref{MC_odd}), which then read
\begin{eqnarray}
  K_6' + \(\nu' - \lam' - \frac{2}{r}\)K_6 - e^{2\lam}K_3 &=& 0\;,\\
  K_3' + \frac{2}{r}K_3 - \frac{l(l+1) - 2}{r^2}K_6 &=& 16\pi(p + \eps)u_3\;.
\end{eqnarray}
These can be easily combined to give a single second-order differential
equation for $K_6$. However, it is more convenient to write this
equation in terms of the variable $h_0 = e^{\nu-\lam} K_6$:
\begin{equation}
  \label{eqh0}
  e^{-2\lam} h_0''-4\pi r(p + \eps)h_0' + \left[8\pi(p + \eps)
    + \frac{4M}{r^3} - \frac{l(l+1)}{r^2}\right]h_0
  = 16\pi e^{\nu-\lam}(p + \eps)u_3\;,
\end{equation}
together with the evolution equation for $u_3$
\begin{equation}\label{equ3}
  \df{}{t}u_3 = -\I m\left[\Omega u_3
  + \frac{2\varpi}{l(l+1)}\(e^{\lam-\nu}h_0 - u_3\)\right]\;.
\end{equation}
At this point, it is worth making some comments on this approximation.
The full set of axial equations (\ref{V4}) -- (\ref{u3}) is a
hyperbolic system describing the propagation of gravitational waves,
which are excited on the one hand by the fluid motion ($r$-modes) and
on the other hand by the curvature of spacetime itself ($w$-modes).
With the above approximation, we have completely suppressed the wave
propagation, and the resulting equations now correspond to a
Newtonian-like picture, where the fluid oscillations are acting as a
source in the equation of the gravitational field. As now being
described by a Poisson-like elliptic equation, implying an infinite
propagation speed, the gravitational field $h_0$ reacts
instantaneously on any changes in the source $u_3$. Of course, this
picture is only an analogy, since the metric variable $h_0$ corresponds
to a post-Newtonian correction of the gravitational field and vanishes
completely in the Newtonian limit.

Furthermore our derivation of this approximation is only valid for
non-barotropic stars.  This is because in general we cannot start from
decoupling the polar and axial equations in the first step as we did.
Instead, when we apply the low-frequency approximation, we actually
have to start from the full coupled system of equations, including
both polar and axial perturbations. If we then do the same ordering in
powers of $\Omega$, we also would have some polar variables of order
$O(1)$, namely the remaining two fluid velocity components, coming
from $\delta u_r$ and the polar part of the angular components
$(\delta u_\theta, \delta u_\phi)$, and the $(rt)$ component of the
metric, usually denoted by $H_1$. It turns out that the polar
constraint equations can be combined to give a single constraint for
$H_1$, which can be reduced to
\begin{equation}
  \label{H1_const}
  \(\Gamma - \Gamma_1\)H_1 = 0\;,
\end{equation}
with 
\begin{equation}
  \Gamma = \frac{p + \eps}{p}\frac{dp}{d\eps}
\end{equation}
the adiabatic index corresponding to the unperturbed configuration and
$\Gamma_1$ the adiabatic index of the perturbed configuration, which
in general differs from $\Gamma$. This is the case for non-barotropic
stellar models, and Eq.~(\ref{H1_const}) can only be satisfied if
$H_1$ vanishes.  But this automatically implies that the polar fluid
perturbations vanish, too, leaving thus only the axial equations,
given above. In the barotropic case, it is $\Gamma = \Gamma_1$ and the
constraint for $H_1$ is trivially satisfied, even for nonzero $H_1$.
But this has as a consequence that the coupling between the polar and
axial mode cannot be neglected, which means that there cannot exist
pure axial mode solutions, since any kind of pure axial initial data
will through the coupling automatically induce polar fluid
oscillations. Hence, our analysis is strictly valid only for
non-barotropic stellar models.

As a further approximation, we could completely neglect all the metric
perturbations. With this so-called relativistic Cowling approximation,
we would be left with a single evolution equation for the fluid
variable $u_3$:
\begin{equation}
  \df{}{t}u_3 = -im\(\Omega - {\frac{2\varpi}{l(l+1)}}\)u_3\;.
\end{equation}
From this equation we can immediately deduce that the various fluid
layers are decoupled from each other, which means that each layer has
its own real oscillation frequency given by
\begin{equation}\label{freqCowl2}
  \sigma = -m\(\Omega - \frac{2\varpi}{l(l+1)}\)\;.
\end{equation}
In the Newtonian limit $(\varpi \rightarrow \Omega)$, this reduces to
the well known result for the frequency of the $r$-mode given in
Eq.~(\ref{freqNewt}).  It should be pointed out that in the
relativistic case, the presence of the frame dragging $\omega$
destroys the occurrence of a single mode frequency and gives rise to a
continuous spectrum, at least to this order of approximation. Of
course, it has been argued that this might be a pure artefact of the
approximation, and the continuous spectrum may disappear as soon as
certain approximations are relaxed.

Let us therefore return to the low-frequency approximation, which is
less restricted than the Cowling approximation, and see whether or not
we can find real mode solutions. To this end, we assume our variables
to have a harmonic time dependence
\begin{eqnarray}
  u_3(t,r) =  u_3(r)e^{-i\sigma t}\;,\\
  h_0(t,r) =  h_0(r)e^{-i\sigma t}\;.
\end{eqnarray}
Note that for the sake of notational simplicity, we do not explicitely
distinguish between the time dependent and time independent form of
the variables. With this ansatz, we assume the $r$-mode frequency
$\sigma$ to be positive for positive values of $m$ in contrast to the
definitions in Eqs.~(\ref{freqCowl2}) and (\ref{freqNewt}). From
Eq.~(\ref{equ3}) we find that
\begin{equation}\label{u3alg}
   u_3 = \frac{2m\varpi}
  {2m\varpi + l(l+1)(\sigma - m\Omega)}e^{\lam-\nu} h_0\;,
\end{equation}
which can be used to eliminate $ u_3$ in Eq.~(\ref{eqh0}), yielding
\begin{equation}
  \label{eveqn}
    \(\sigma - m\Omega + \frac{2m\varpi}{l(l+1)}\)
    \left[e^{-2\lam} h_0''-4\pi r(p + \eps)h_0'
    -\(8\pi(p + \eps) - \frac{4M}{r^3}
      + \frac{l(l+1)}{r^2}\)h_0\right]
    + 16\pi(p + \eps)(\sigma - m\Omega)h_0 = 0\;.
\end{equation}
With appropriate boundary conditions, this equation represents an
eigenvalue problem which should yield one, or possibly several,
distinct eigenmodes. However, as was at first pointed out by Kojima
(1997,1998), it might occur that the denominator in Eq.~(\ref{u3alg})
can become zero at some point inside the star, and the resulting
eigenvalue problem becomes singular at this point. If the zero of the
denominator lies outside the star, the eigenvalue problem is regular,
since outside $u_3 = 0$ and Eq.~(\ref{eqh0}) can be directly solved
without using Eq.~(\ref{u3alg}). The zeroes of the denominator occur
if the frequency $\sigma$ lies in an interval determined by the values
of $\varpi$ at the centre and the surface, which we denote by
$\varpi_c$ and $\varpi_s$, respectively:
\begin{equation}\label{range}
   m\Omega\(1 - \frac{2\varpi_s}{\Omega l(l+1)}\) < \sigma
  < m\Omega\(1 - \frac{2\varpi_c}{\Omega l(l+1)}\)\;.
\end{equation}
By comparison with similar results from fluid dynamics, Kojima
concluded that this equation should have a continuous spectrum with
the frequency range given by (\ref{range}). This was put on a rigorous
mathematical footing by Beyer \& Kokkotas (1999). However, they could
not exclude that there might not exist additional isolated
eigenvalues, which would correspond to true mode solutions.

We will now show that there actually exist such solutions, even though
in some cases they are unphysical since the corresponding fluid
perturbations would be divergent at the singular point. To make things
look simpler, we can rescale Eq.~(\ref{eveqn}) and make it independent
of $\Omega$ and $m$. Following Lockitch et al.~(2001), we introduce a
normalized frequency
\begin{equation}
  \alpha = \half l(l+1)\(1 - \frac{\sigma}{m\Omega}\)
\end{equation}
and rewrite Eq.~(\ref{eveqn}) as
\begin{equation}\label{eveqn_a}
  \(\alpha - \hat\varpi\)
  \left[e^{-2\lam} h_0''
    -4\pi r(p + \eps)h_0' -\(8\pi(p + \eps) - \frac{4M}{r^3}
    + \frac{l(l+1)}{r^2}\)h_0\right] + 16\pi(p + \eps)\alpha h_0 = 0\;,
\end{equation}
where 
\begin{equation}
  \hat\varpi := \varpi/\Omega\;.
\end{equation}
Equation (\ref{u3alg}) then reads
\begin{equation}\label{u3alg_a}
   u_3 = \frac{\hat\varpi}
  {\hat\varpi - \alpha}e^{\lam-\nu} h_0\;.
\end{equation}
Equation (\ref{eveqn_a}) becomes singular if $ \alpha$ lies in the interval
limited by the values of $\hat\varpi$ at the centre and at the
surface of the star, i.e.~if
\begin{equation}
  \hat\varpi_c < \alpha < \hat\varpi_s\;.
\end{equation}
For a solution to be acceptable, it must be regular at the origin,
which amounts to $h_0(0) = 0$, and it must vanish at infinity. As
already mentioned above the integration of Eq.~(\ref{eveqn}) is
straightforward, if the singular point lies outside the star. It is
only when the singular point lies inside the star that some care has
to be taken.

Let us now assume the singular point $r = r_0$ lie inside the star.
An analysis of Eq.~(\ref{eveqn}) (Andersson 1998b) shows that the
singular point is a regular singularity, which admits a Frobenius
expansion of the form
\begin{eqnarray}
  \label{expan}
  h_0(r - r_0) &=&  A \(a_1 (r - r_0) + a_2 (r - r_0)^2 + \ldots \) + \non\\
  && B \left[\(a_1 (r - r_0) + a_2 (r - r_0)^2 + \ldots\)\ln|r - r_0|
    + b_0  + b_1(r - r_0) + b_2(r - r_0)^2 + \ldots \right]\;.
\end{eqnarray}
Even though the solution is finite and smooth at the singular point $r
= r_0$, its derivative diverges, because of the logarithmic term.
Moreover, if we want to compute the associated velocity component $
u_3$, we find that unless $h_0(r = r_0) = 0$, it will blow up at the
singular point $r = r_0$. Therefore, we conclude that the coefficient
$B$ has to vanish in order to obtain a physical solution and we are
only left with the first power series starting with the linear term
$a_1(r - r_0)$.  This yields vanishing $h_0$ at $r = r_0$ and
therefore $u_3$ can be finite at this point.  The question is whether
there are solutions satisfying both $h_0(r = r_0) = 0$ and the
appropriate boundary conditions at the centre and at infinity. We will
now show that this cannot be the case.

Suppose that $h_0(r = r_0) = 0$ and $h_0'(r = r_0) > 0$.  We can then
integrate Eq.~(\ref{eveqn}) from $r_0$ to $r > r_0$:
\begin{equation}
  h_0'(r) = h_0'(r_0) + \int_{r_0}^r{e^{2\lam}\left[4\pi r(p + \eps)h_0'
  +\(8\pi(p + \eps) - \frac{4M}{r^3} + \frac{l(l+1)}{r^2}\)h_0
  + 16\pi(p + \eps)\frac{\alpha}{\hat\varpi - \alpha} h_0\right]dr},
\end{equation}
Since it is $\varpi - \alpha > 0$ for $r > r_0$, each coefficient in
the integral is strictly positive, hence we will have $h_0'(r) > 0$
for all $r > r_0$; i.e.~$h_0$ keeps increasing as $r \rightarrow
\infty$, which is clearly incompatible with our requirement that $h_0$
vanish at infinity. Of course, the same argument holds if $h_0' < 0$
at $r = r_0$, with $h_0$ keeping decreasing. Hence, it follows that
$h_0(r = r_0) \ne 0$, but this means that we cannot have a vanishing
coefficient $B$. Therefore, our solution will always contain the
logarithmic term, which means that the associated velocity
perturbation $u_3$ is divergent at this point. This is clearly
unphysical. We thus conclude that it is in principle possible to find
eigensolutions to Eq.~(\ref{eveqn_a}), however, if the associated
eigenfrequencies lie inside the continuous band, the solutions become
singular and have to be discarded on physical grounds.

It has been shown by Lockitch et al.~(2001) that for the existence of
mode solutions, the allowed range of the eigenvalues $\alpha$ is
bounded from below by $\hat\varpi_c$:
\begin{equation}
  \label{bound1}
   \hat\varpi_c \le \alpha \le 1\;.
\end{equation}
However, based on our above argumentation, we can further restrict this
interval for the physically allowed eigenmodes to have as lower limit
the value of $\hat\varpi_s$:
\begin{equation}
  \label{bound2}
   \hat\varpi_s < \alpha \le 1\;.
\end{equation}

\section{Numerical results}

The numerical integration of Eq.~(\ref{eveqn_a}) can be easily
accomplished if the singular point lies outside the star, since in the
exterior $u_3 = 0$ and we can therefore use the non-singular
Eq.~(\ref{eqh0}) for the integration toward the outer boundary. If the
singular point lies inside the star, we initiate our integration with
a regular solution at the origin and integrate outward close to the
singular point, where we match the solution to the expansion
(\ref{expan}), i.e.~we compute the leading coefficients $b_0$ and
$a_1$. This gives us the new starting values to the right of the
singular point and we can continue the integration up to a finite
point outside the star, where we test if the solution satisfies the
correct boundary condition. We mention again that for the integration
outside the star we take the non-singular Eq.~(\ref{eqh0}), with $u_3$
set to zero.

\begin{figure}
\begin{center}
\vspace*{1cm}
\epsfxsize=8cm
\epsfbox{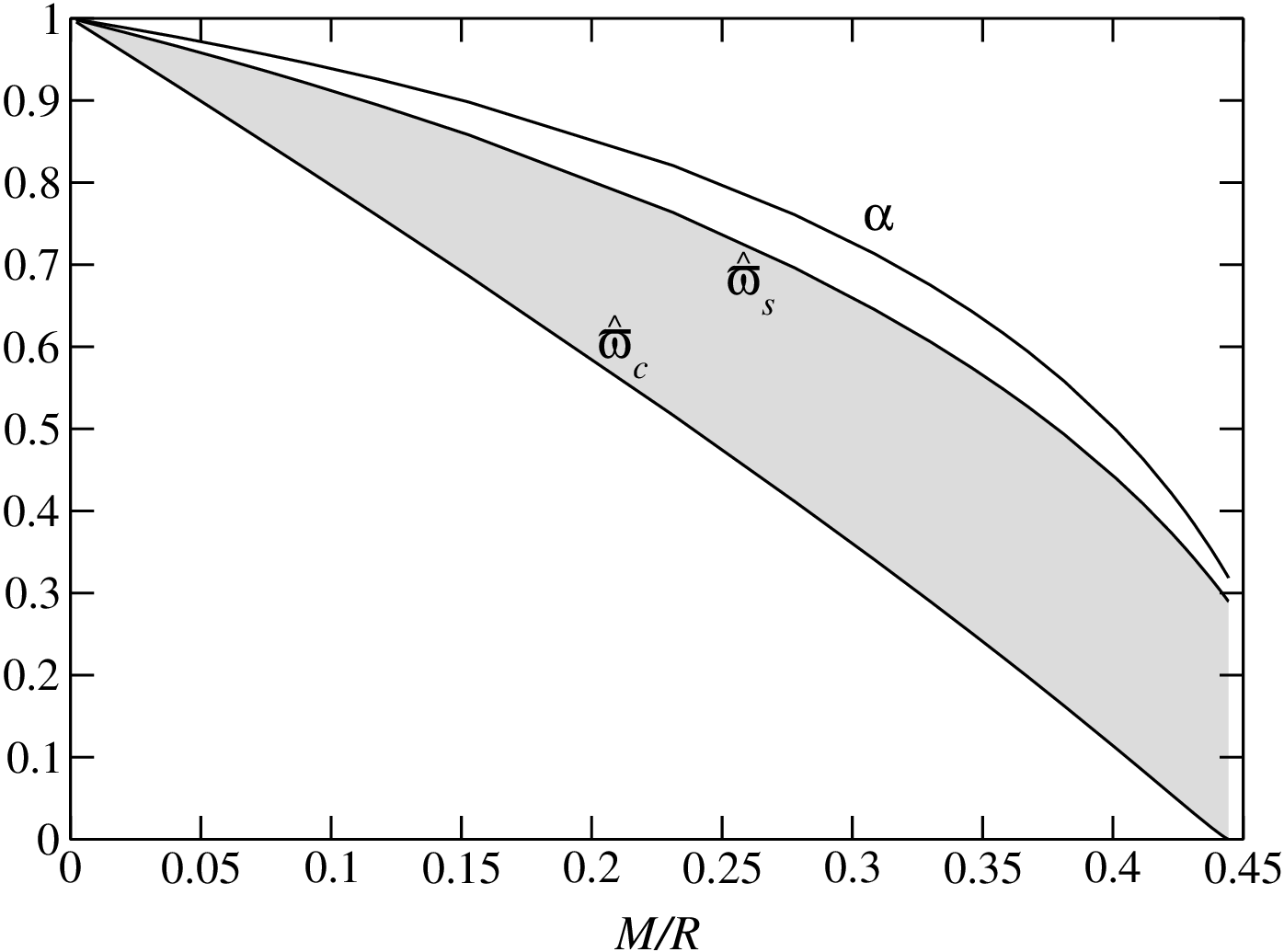}
\vspace*{5mm}
\caption{\label{fig:em_cd}
  Boundaries of the continuous spectrum $\hat\varpi_c$ and
  $\hat\varpi_s$ together with the $r$-mode frequency $\alpha$ as a
  function of compactness $M/R$ for uniform density models. The mode
  frequency $\alpha$ lies always outside the continuous band (shaded
  area).}
\end{center}
\end{figure}

\begin{figure}
\begin{center}
\vspace*{1cm}
\leavevmode
\epsfxsize=8cm
\epsfbox{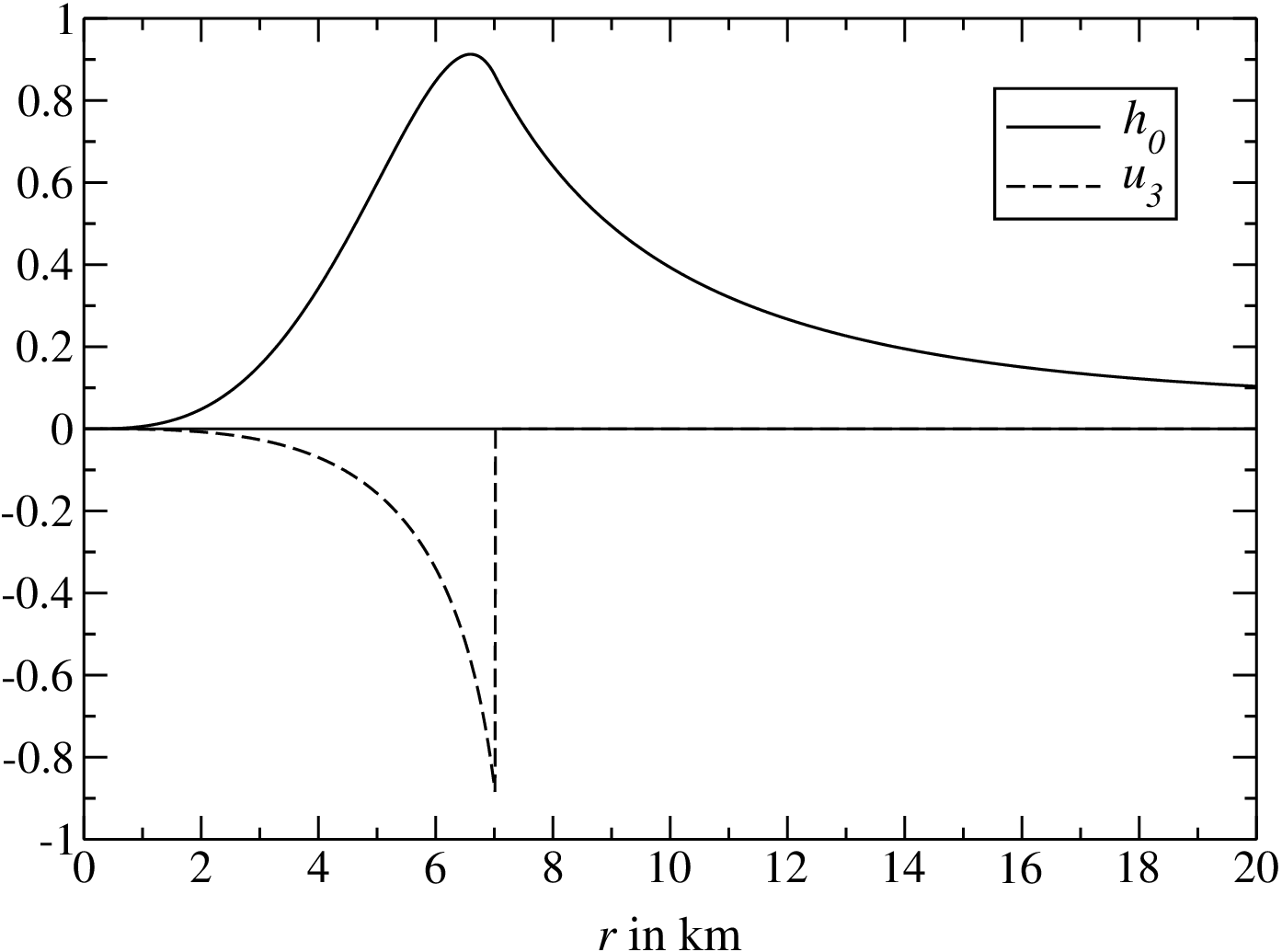}
\vspace*{5mm}
\caption{\label{fig:ef_cd}
  Eigenfunctions $h_0$ and $u_3$ for a uniform density model with
  compactness $M/R = 0.153$. The corresponding $r$-mode frequency is
  $\alpha = 0.89806$.}
\end{center}
\end{figure}

We have performed mode calculations for sequences of uniform density
and polytropic stars. For the uniform density models, the
eigenfrequency $\alpha$ always lies outside the range of the
continuous spectrum and therefore the associated eigenfunctions do not
exhibit any singularities. In Fig.~\ref{fig:em_cd}, we show the
normalized eigenvalues $\alpha$ for $l = 2$ as a function of the
compactness $M/R$ together with $\hat\varpi_c$ and $\hat\varpi_s$,
marking the boundaries of the continuous spectrum. For larger $l$ (not
shown), the eigenvalues $\alpha$ decrease and converge to
$\hat\varpi_s$, but stay always above $\hat\varpi_s$. \footnote{ Note
  that in Table 1 of Lockitch et al.~(2001), there is a systematic
  error in their given values of $\alpha$, which are too large by
  about 5 per cent.  This might be a consequence of a misprint in
  their Eqs.~(5.2), (5.4) and (5.7), where the terms $\(1 -
  2M_0/R\)^{1/2}$ and $\(1 - 2M_0/R \(r/R\)^2\)^{1/2}$ got confused.}
In Fig.~\ref{fig:ef_cd}, we show the eigenfunctions $h_0$ and $u_3$
for a uniform density model with a compactness of $M/R = 0.153$ and
corresponding mode frequency $\alpha = 0.89806$. Close to the centre
of the star, the fluid perturbation $u_3$ is proportional to
$r^{l+1}$, but as it approaches the stellar surface, it grows much
stronger, which comes from the denominator in Eq.~(\ref{u3alg_a})
becoming very small.

\subsection{Polytropic models}

\begin{figure}
\begin{center}
\vspace*{1cm}
\leavevmode
\epsfxsize=8cm
\epsfbox{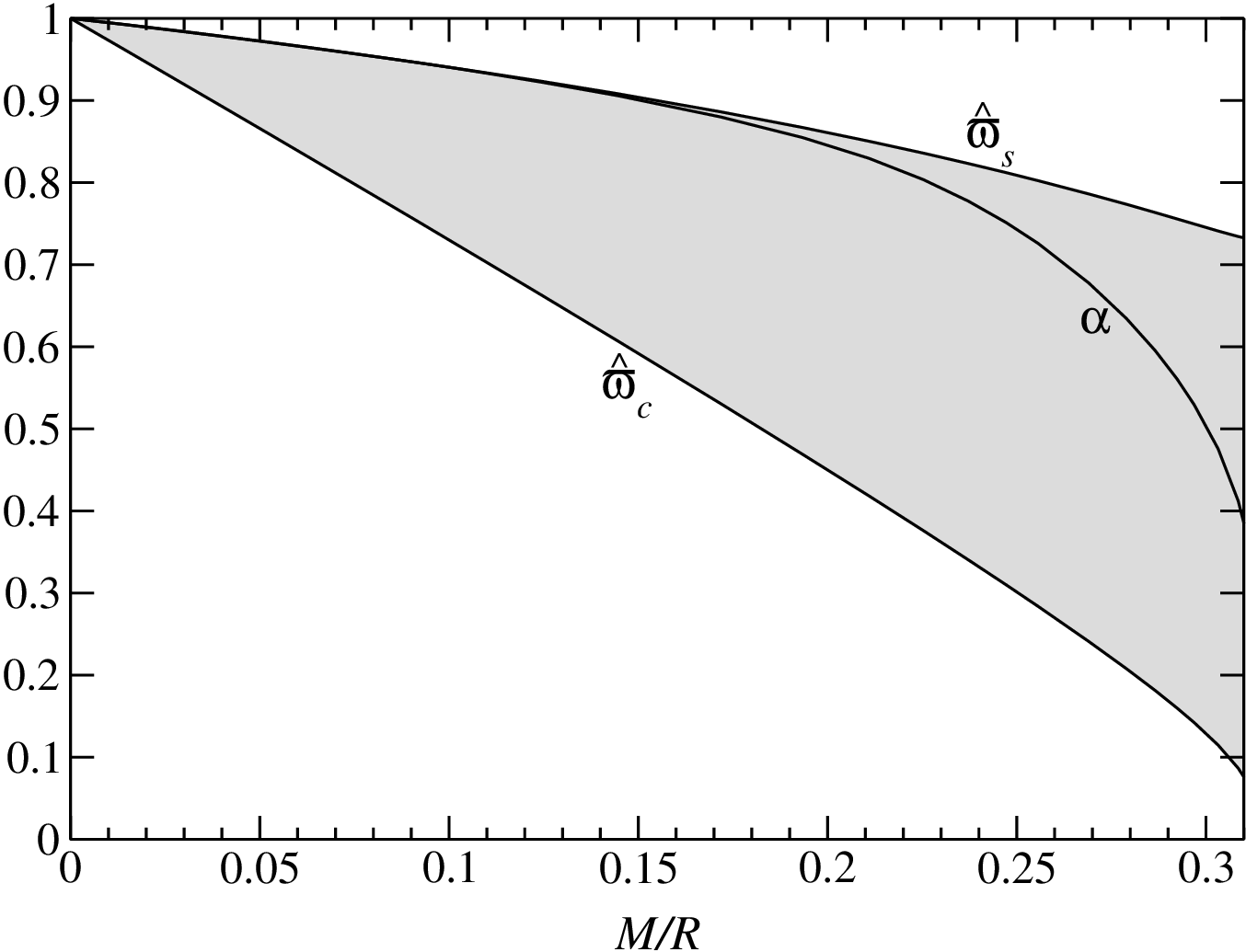}
\vspace*{5mm}
\caption{\label{fig:em_poly}
  Boundaries of the continuous spectrum $\hat\varpi_c$ and
  $\hat\varpi_s$ together with the $r$-mode frequency $\alpha$ as a
  function of compactness $M/R$ for polytropic $n = 1$ models. For
  $M/R < 0.085$, the mode frequency $\alpha$ lies outside the
  continuous band (shaded area), but differs from the value of
  $\hat\varpi_s$ not more than 0.01 per cent. For $M/R > 0.085$ it
  migrates inside the band.}
\end{center}
\end{figure}

For polytropic models, obeying an equation of state of the form
\begin{equation}
  p = \kappa\eps^{1 + 1/n}
\end{equation}
with polytropic index $n$, we obtain a quite different picture. For a
polytropic index $n = 1$, as it is for instance shown in
Fig.~\ref{fig:em_poly}, it is only for the less compact stellar models
that the eigenfrequencies $\alpha$ lie outside the continuous spectrum
and therefore represent physical mode solutions.  However, they are
already that close to the upper boundary of the continuous spectrum
$\hat\varpi_s$ that in Fig.~\ref{fig:em_poly} they cannot be
distinguished any more. For more compact models, the eigenfrequency
eventually moves inside the domain of the continuous spectrum, which
means that the singular point now lies inside the star.  This happens
for a compactness of about $M/R \approx 0.085$. As discussed above, at
the singular point the mode solution for $h_0$ exhibits an infinite
slope and the associated fluid perturbation $u_3$ diverges. Therefore,
we have to discard them as being unphysical mode solutions. In Table
\ref{models}, we have listed some polytropic models with their
physical parameters and the eigenvalues $\alpha$ for $l=2$ and $l=3$.
The frequencies which are marked by an asterisk lie inside the
continuous band and therefore correspond to unphysical mode solutions.
For $l=2$, only models 1 and 2 permit physical modes, whereas for
$l=3$, the modes are unphysical for all the stellar models. We should
also mention that all our values are in perfect agreement with those
previously obtained by Andersson (1998b).

\begin{figure}
\begin{center}
\vspace*{1cm}
\leavevmode
\epsfxsize=8cm
\epsfbox{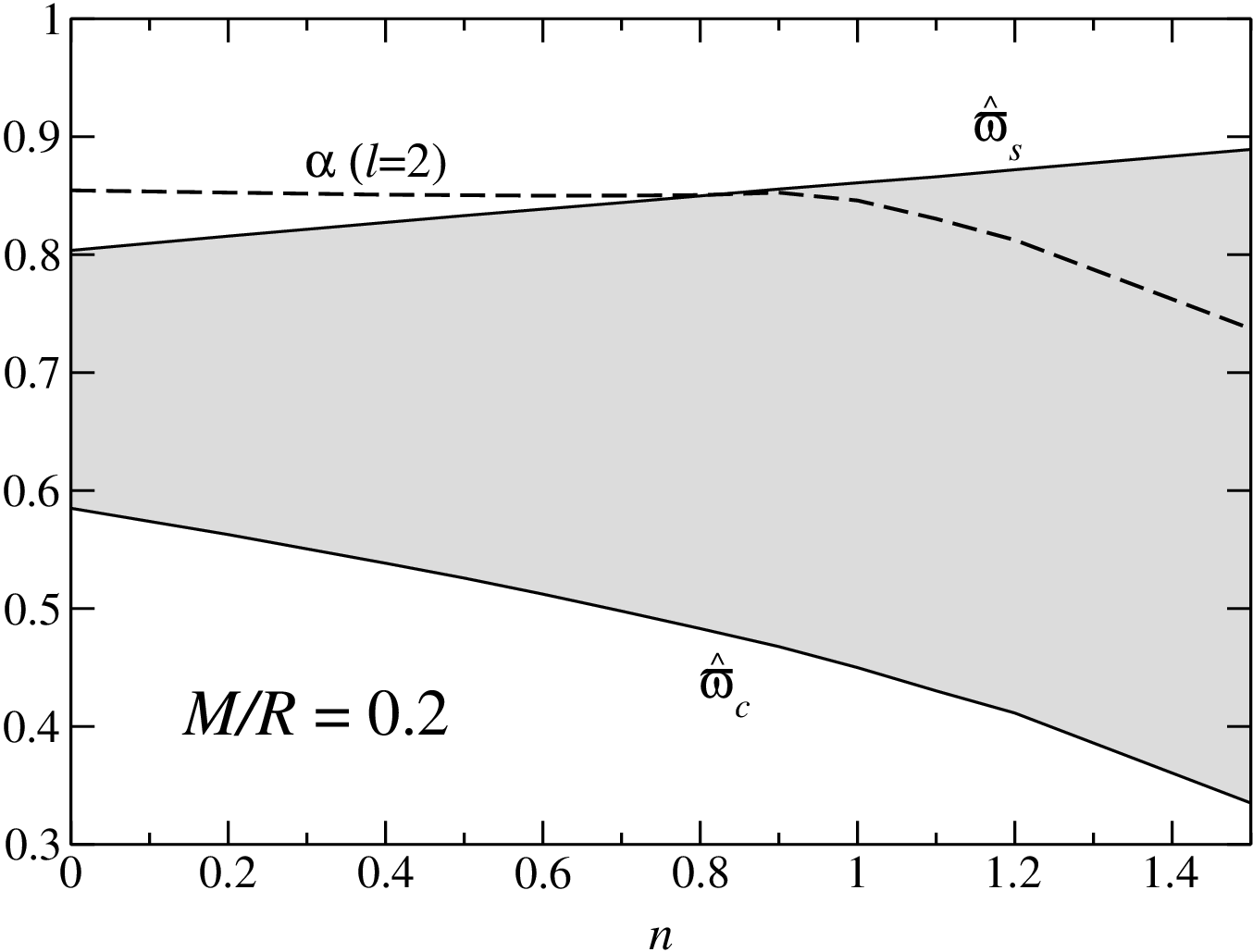}
\vspace*{5mm}
\caption{\label{fig:index}
  The boundaries of the continuous spectrum $\hat\varpi_c$ and
  $\hat\varpi_s$ together with the $r$-mode frequency $\alpha$ as a
  function of the polytropic index $n$ for stellar models with
  compactness of $M/R = 0.2$.  The case $n=0$ corresponds to a
  uniform density model.}
\end{center}
\end{figure}

To assess how the existence of a physical mode solution depends on the
polytropic index $n$, we have computed modes for stellar models with
fixed compactness $M/R$ but with different values of $n$, ranging from
0 to 1.5. The results are depicted in Fig.~\ref{fig:index}, where we
show $\alpha$ as a function of $n$ for $l=2$. For small values of $n$,
i.e.~for stiff equations of state, the mode eigenfrequency lies
outside the range of the continuous spectrum. But as $n$ is increased,
what corresponds to softening the equation of state, the mode
frequency eventually crosses the boundary and migrates inside the
continuous spectrum. This happens at $n \approx 0.8$, but for larger
values of $l$, the transition point moves to smaller values of $n$.
Actually, it is not the mode frequency $\alpha$ which moves towards
the boundary of the continuous spectrum $\hat\varpi_s$ as $n$ is
increased, it is rather the boundaries of the continuous spectrum
which start to expand, and $\hat\varpi_s$ approaches the mode
frequency $\alpha$, which more or less hovers at a constant value. For
$n = 0$, the uniform density models, the range of the continuous
spectrum (the shaded area in Fig.~\ref{fig:index}) is the smallest,
and probably it is only in this case that one can find eigenvalues for
quite large $l$, if not for all $l$. We should also mention that for
each polytropic model, there seems to exist a maximal value of $l$,
beyond which there are neither physical nor unphysical mode solutions.
The frequency $\alpha$ of the unphysical mode solution quickly
approaches $\hat\varpi_c$ as $l$ is continuously increased.  For $l$
larger than the critical value, where $\alpha(l) = \hat\varpi_c$, we
could not find any mode solution at all. For the $n=1$ models of Table
1, this happens already for $l=4$.

\begin{figure}
\begin{center}
\vspace*{1cm}
\leavevmode
\begin{minipage}{8cm} 
\epsfxsize=\textwidth
\epsfbox{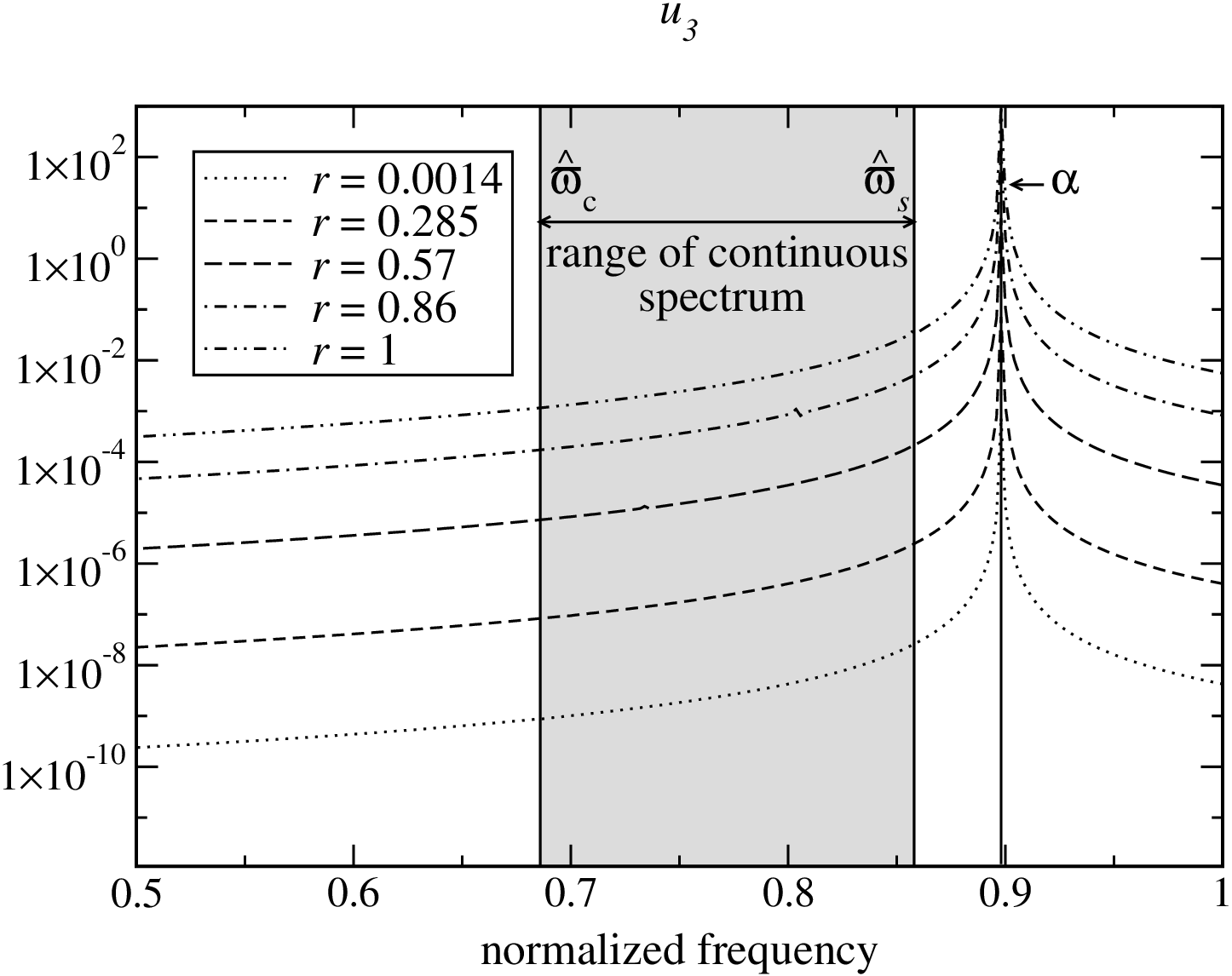}
\end{minipage}
\hspace*{5mm}
\begin{minipage}{8cm}
\epsfxsize=\textwidth
\epsfbox{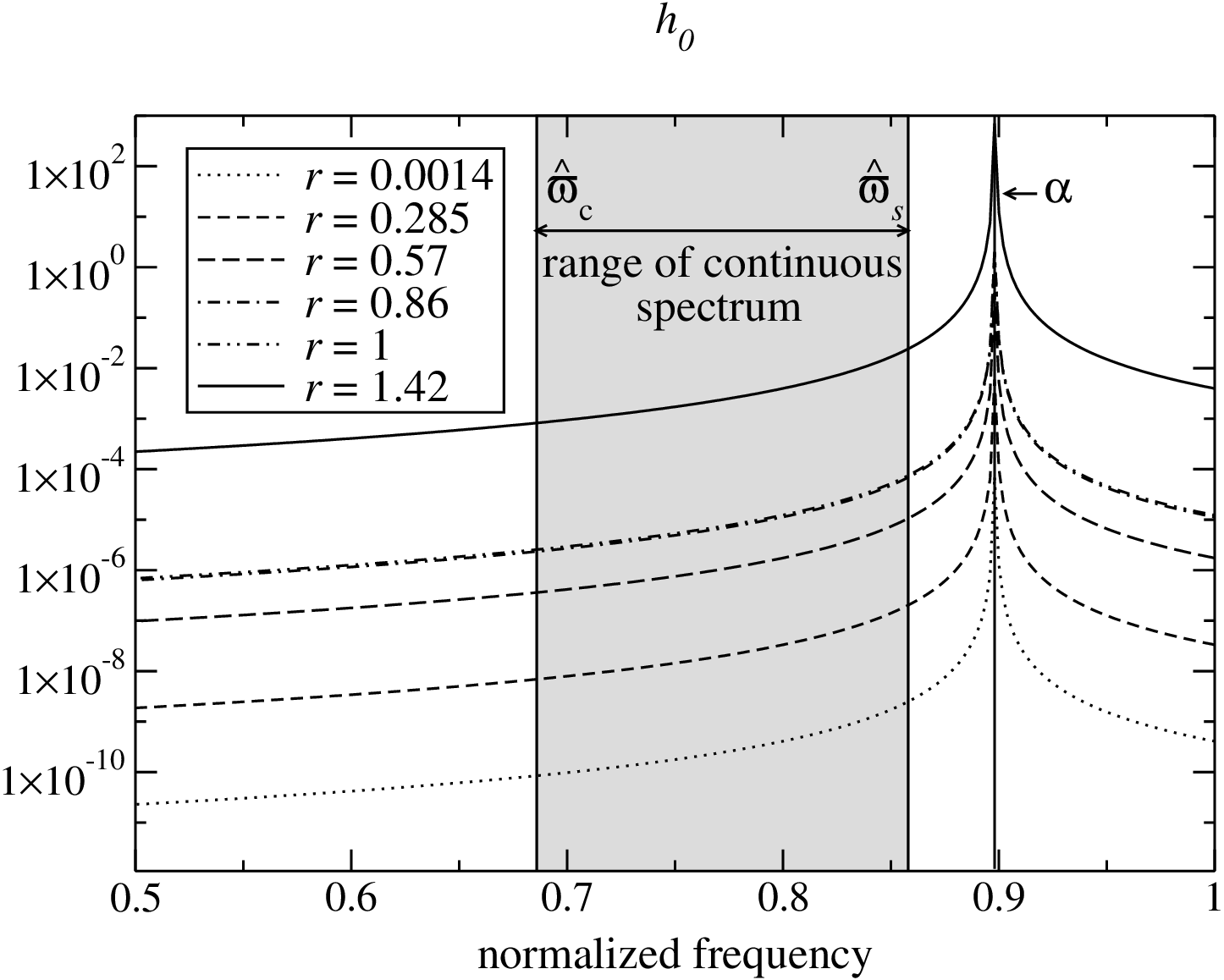}
\end{minipage}
\vspace*{5mm}
\caption{\label{fig:ps_cd_mode}
  Power spectrum of the time evolution of a mode solution (initial
  data of Fig.~\ref{fig:ef_cd}) for the uniform density model. For
  both the fluid $u_3$ (left graph) and the metric variable $h_0$
  (right graph), the spectrum shows a single peak at the expected
  frequency of $\alpha = 0.89806$. In each graph, the power spectrum
  has been taken at 5 different locations inside the star. For $h_0$,
  there is an additional one from outside the star.  The spectra have
  been individually rescaled in order to give clearer graphs.}
\end{center}
\end{figure}

To check and corroborate our above results, we also numerically
evolved the time dependent equations (\ref{eqh0}) and (\ref{equ3}) and
took Fourier transforms of the resulting evolution. For initial data
representing the physical mode solution of Fig.~\ref{fig:ef_cd}, the
time evolution indeed gives a single frequency signal at each point
inside and outside the star. In this case, there is no sign of a
continuous spectrum at all, and all the fluid layers oscillate in a
uniform manner. This is shown in Fig.~\ref{fig:ps_cd_mode}, where for
both the fluid variable $u_3$ (left panel) and the metric variable
$h_0$ (right panel), there is one single peak, which is independent of
the location $r$.

\begin{figure}
\begin{center}
\vspace*{1cm}
\leavevmode
\epsfxsize=8cm
\epsfbox{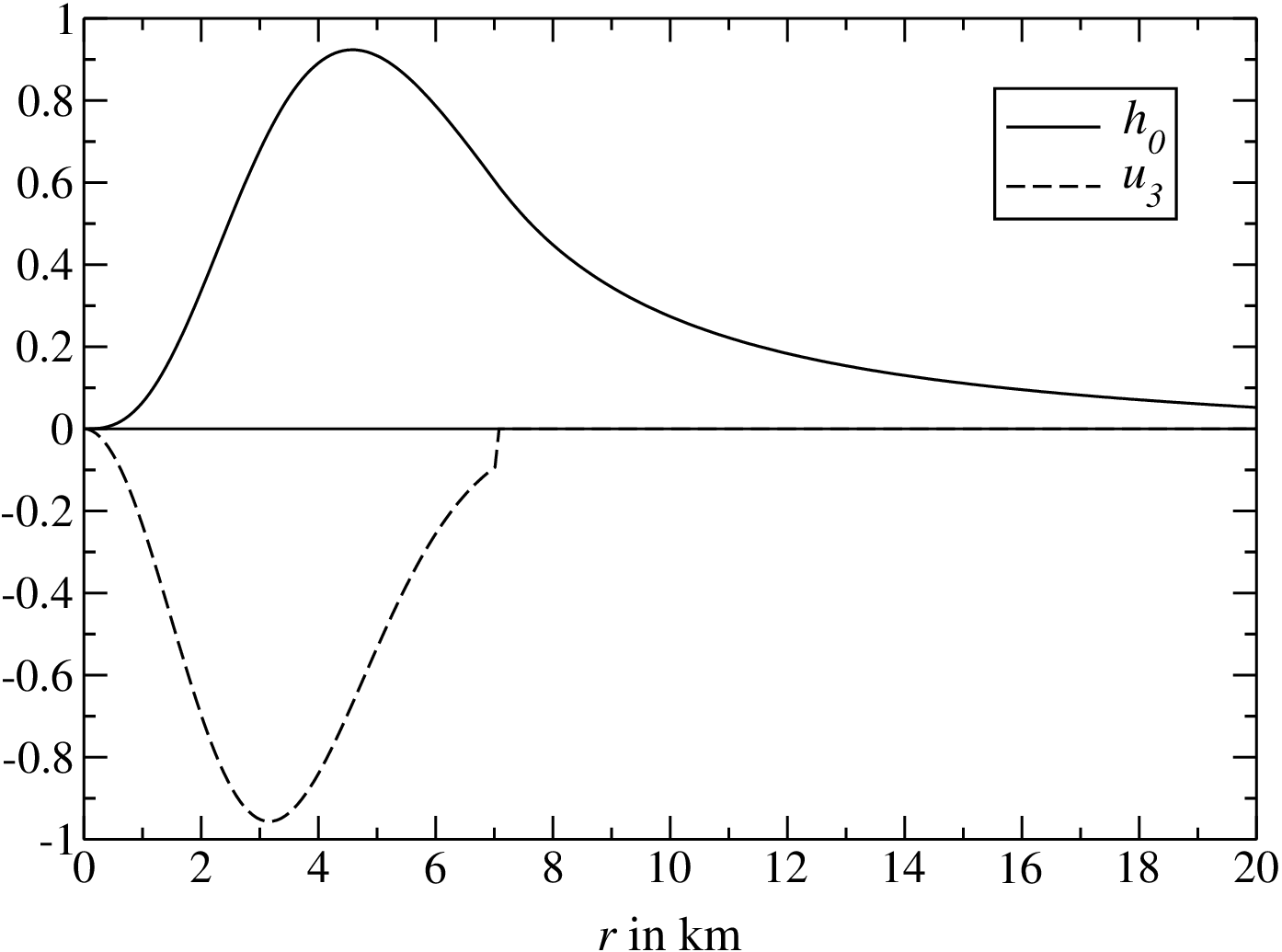}
\vspace*{5mm}
\caption{\label{fig:arb_id}Arbitrary initial data for the uniform
density model.}
\end{center}
\end{figure}

\begin{figure}
\begin{center}
\vspace*{1cm}
\leavevmode
\begin{minipage}{8cm} 
\epsfxsize=\textwidth
\epsfbox{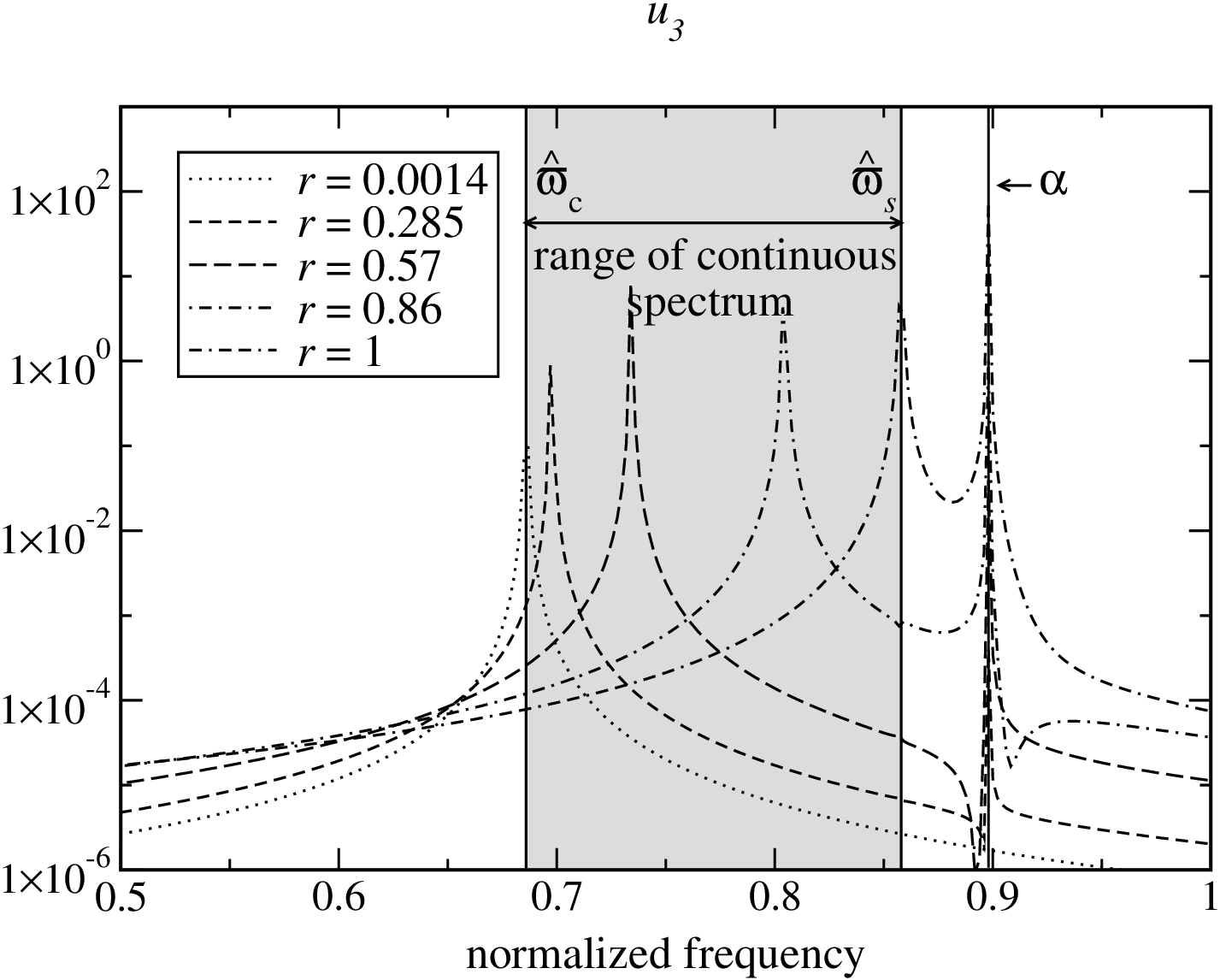}
\end{minipage}
\hspace*{5mm}
\begin{minipage}{8cm}
\epsfxsize=\textwidth
\epsfbox{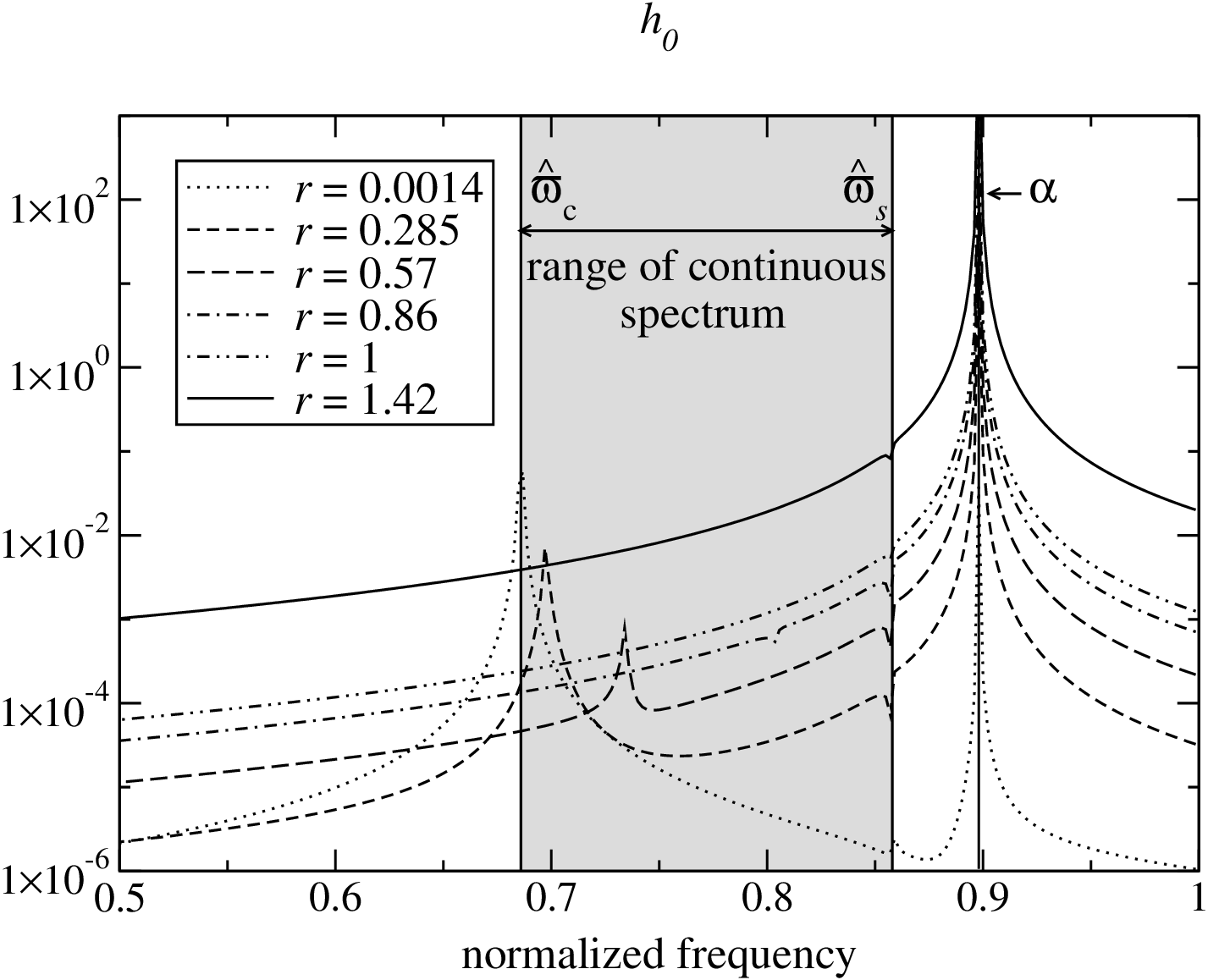}
\end{minipage}
\vspace*{5mm}
\caption{\label{fig:ps_cd_arb}
  Late time power spectrum of the time evolution of arbitrary initial
  data (initial data of Fig.~\ref{fig:arb_id}) for the uniform
  density model. At each location $r$ inside the star, the fluid $u_3$
  (left graph) shows two peaks, one corresponding to the $r$-mode
  frequency $\alpha = 0.89806$ and another corresponding to the value
  of $\hat\varpi$ at this particular location $r$. The spectrum of the
  metric variable $h_0$ (right graph) is quite similar, however, for
  the spectra corresponding to locations close to the surface and
  outside the star, the influence of the continuous spectrum becomes
  negligible and the mode dominates. Again, the amplitudes of the
  spectra are arbitrarily rescaled.}
\end{center}
\end{figure}

If we now choose arbitrary initial data, as for instance in
Fig.~\ref{fig:arb_id}, we expect the power spectrum at a given
location $r$ to consist of two peaks: One which is independent of the
location inside the star and represents the eigenmode, and another
peak which varies between the boundaries determined by $\hat\varpi_c$
and $\hat\varpi_s$ as one moves throughout the star.  This is how the
continuous spectrum should show up in the Fourier transform. For the
fluid variable $u_3$, the power spectrum of the evolution indeed
confirms our expectations, as is shown in the left panel of
Fig.~\ref{fig:ps_cd_arb}.

However, the spectra of $h_0$ show that for locations closer to the
stellar surface, the peaks corresponding to the continuous spectrum
are smaller by several orders of magnitudes compared to the peak
representing the eigenmode. For $u_3$, the peaks are of the same order
of magnitude. And outside the star, $h_0$ shows only the eigenmode
peak, and no sign of the presence of the continuous spectrum, which
should reveal itself as a superposition of all the frequencies in the
range between $\hat\varpi_c$ and $\hat\varpi_s$.  In therefore seems
to be invisible for an external observer. We should note that those
spectra are taken after a certain initial time, in which the system
adjusts itself. If we had taken the Fourier transform right from
$t=0$, we would have obtained a clear sign of the continuous spectrum.

Let us now turn or attention to the polytropic cases, where we can
have unphysical mode solutions. We will present evolution runs for the
stellar models 1 and 5 from Table \ref{models} with $l=2$. For model
1, the singular point lies outside the stellar surface and therefore
there exists a physical mode solution. For model 5, the singular point
lies inside the star, hence $\alpha < \hat\varpi_s$ and the associated
eigensolution is unphysical. It should be noted that this model is
also unstable with respect to radial collapse.

\begin{table}
\caption{\label{models}List of polytropic stellar models with
$n = 1$ and $\kappa = 100\,$km$^2$.}
\begin{tabular}{ccccccccc}
\rule[-2.5mm]{0mm}{7.5mm}
Model & $\eps_0\;[$g/cm$^3]$ & $M\;[M_\odot]$ & $R\;[$km$]$ & $M/R$
& $\hat\varpi_c$ & $\hat\varpi_s$ & $\alpha(l=2)$ & $\alpha(l=3)$ \\
\hline
\rule[0mm]{0mm}{5mm}
1 & $1.0\times 10^{14}$ & 0.120 & 12.32 & 0.014 & 0.96168 & 0.99237
& 0.99254 & 0.98523$^*$ \\
2 & $5.0\times 10^{14}$ & 0.495 & 11.58 & 0.063 & 0.83048 & 0.96431
& 0.96453 & 0.92446$^*$ \\
3 & $1.0\times 10^{15}$ & 0.802 & 10.81 & 0.109 & 0.70420 & 0.93407 
& 0.93362$^*$ & 0.84895$^*$ \\
4 & $5.0\times 10^{15}$ & 1.348 & 7.787 & 0.256 & 0.28377 & 0.80236
& 0.72579$^*$ & 0.44782$^*$ \\
5 & $1.0\times 10^{16}$ & 1.300 & 6.466 & 0.297 & 0.14214 & 0.74960
& 0.52932$^*$ & 0.25301$^*$ \\
\end{tabular}
\end{table}

\begin{figure}
\begin{center}
\vspace*{1cm}
\leavevmode
\epsfxsize=8cm
\epsfbox{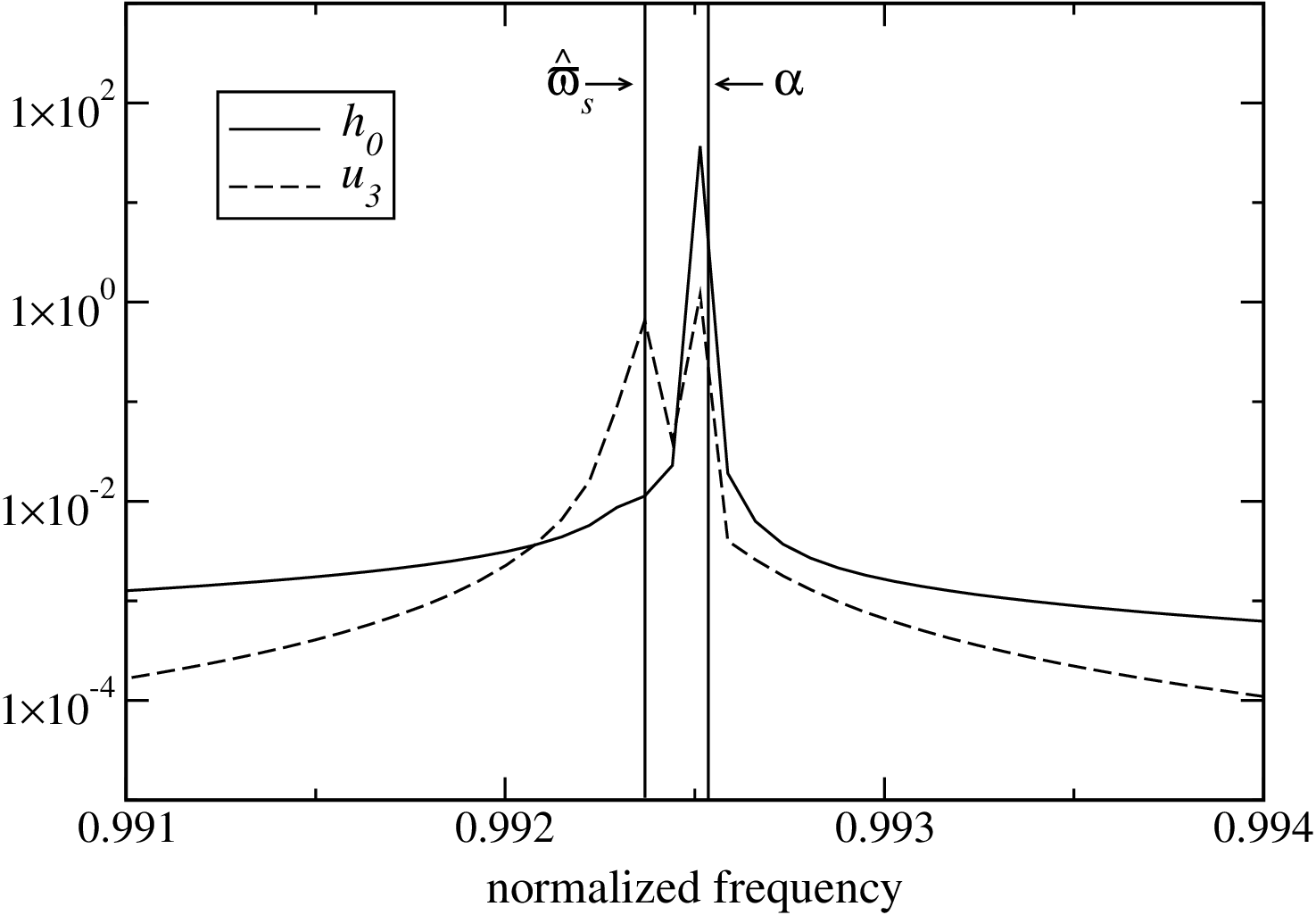}
\vspace*{5mm}
\caption{\label{fig:ps_poly_1e14}
  Power spectrum of the time evolution of arbitrary initial data for
  the polytropic model 1, taken at the surface. Both $u_3$ and $h_0$
  show a peak at the expected $r$-mode frequency $\alpha = 0.99254$.
  But only $u_3$ shows an additional peak at $\hat\varpi =
  \hat\varpi_s = 0.99237$.}
\end{center}
\end{figure}

For model 1, the physical mode solution can be used as initial data.
As for the uniform density case, the numerical evolution of such data
yields a purely sinusoidal oscillation with the expected $r$-mode
frequency $\alpha$. Therefore, the corresponding power spectrum is
similar to Fig.~\ref{fig:ps_cd_mode}.  For arbitrary initial data, we
obtain a picture similar to Fig.~\ref{fig:ps_cd_arb}. Note, that the
values of $\alpha$ and $\hat\varpi_s$ differ only by about 0.01 per
cent. Still, with a high resolution run we can numerically distinguish
these values, as is shown in Fig.~\ref{fig:ps_poly_1e14}. Here, we
plot the power spectra of $h_0(t)$ and $u_3(t)$ taken at the stellar
surface. The spectrum of $h_0$ shows one single peak at the
eigenfrequency $\alpha$, whereas $u_3$ shows two peaks at $\alpha$ and
$\hat\varpi_s$.

\begin{figure}
\begin{center}
\vspace*{1cm}
\leavevmode
\begin{minipage}{8cm} 
\epsfxsize=\textwidth
\epsfbox{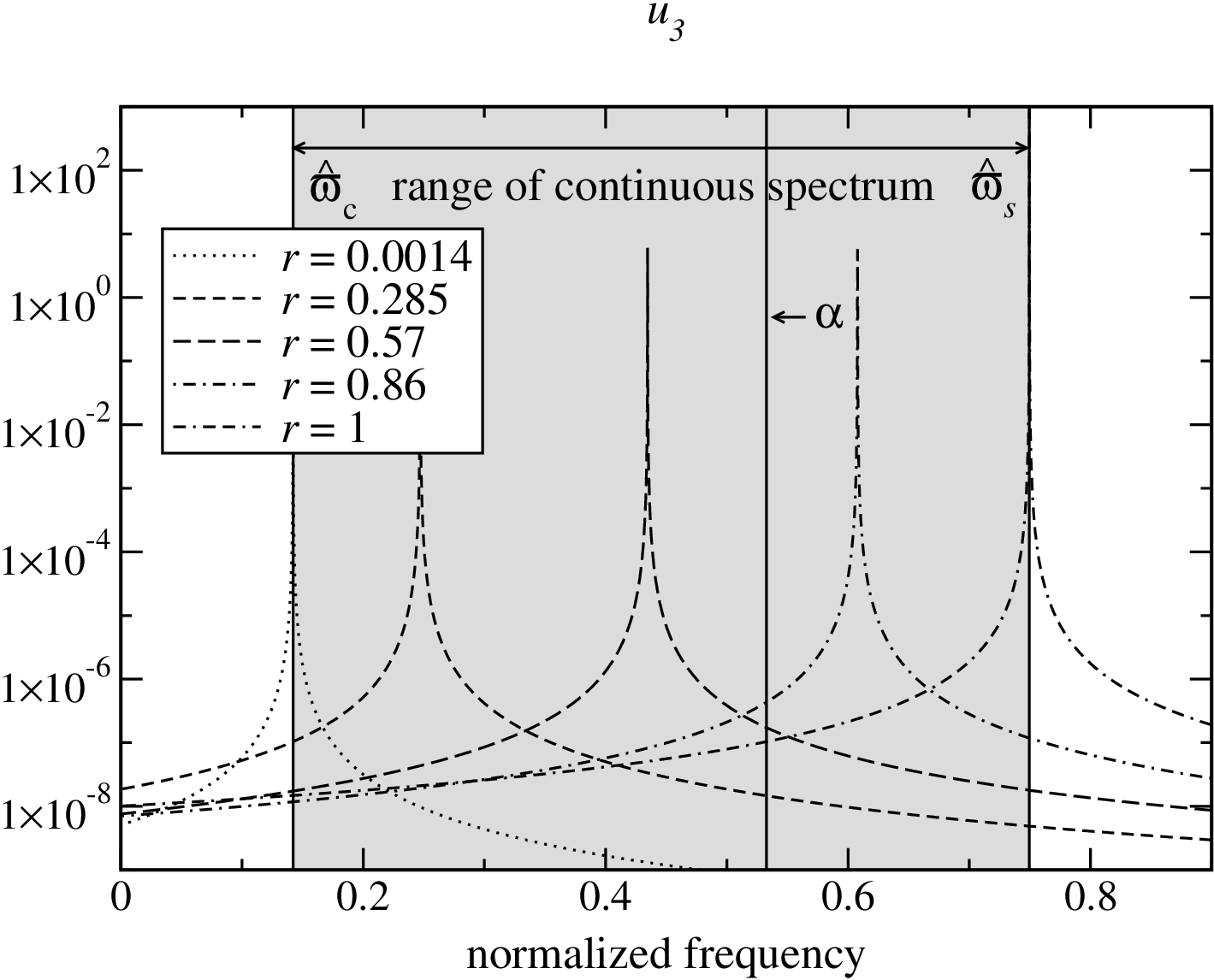}
\end{minipage}
\hspace*{5mm}
\begin{minipage}{8cm}
\epsfxsize=\textwidth
\epsfbox{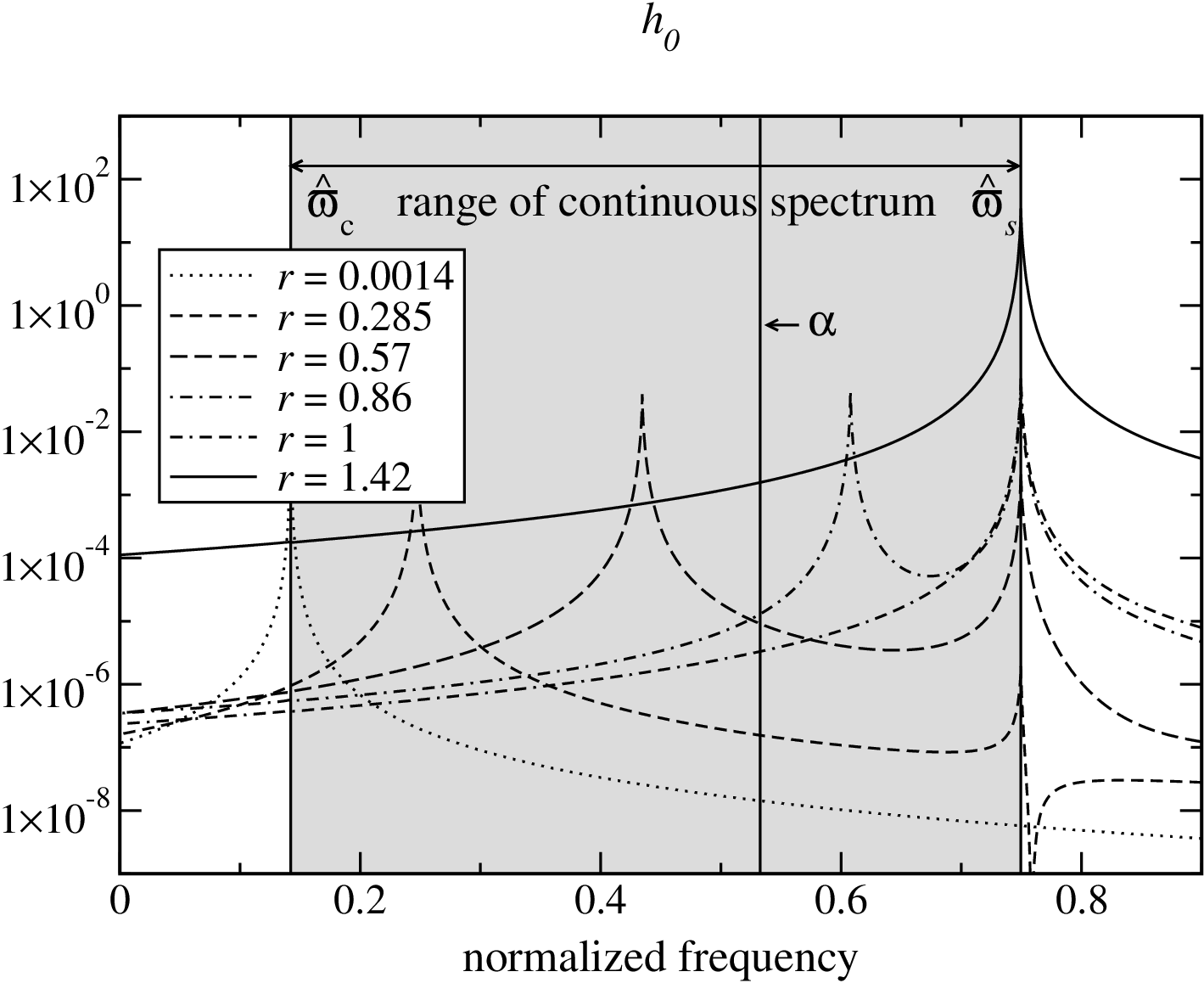}
\end{minipage}
\vspace*{5mm}
\caption{\label{fig:ps_poly_1e16}
  Late time power spectrum of the time evolution of arbitrary initial
  data for the polytropic model 5. For this model, we do not expect a
  physical mode to exist. The frequency of the unphysical mode is
  given by $\alpha = 0.52932$. It is clear that at this location,
  neither $u_3$ nor $h_0$ show a peak in the power spectra. At each
  location, the spectrum of $u_3$ shows a single peak corresponding to
  the respective value of $\hat\varpi$. However, in addition to those
  variable peaks, the spectra of $h_0$ reveal a common peak with the
  frequency given by $\hat\varpi_s$. This peak can be traced back to
  the kink in the energy density profile at the surface.}
\end{center}
\end{figure}

For model 5, things are quite different. Here, we cannot evolve
initial data representing the unphysical mode solutions because the
fluid perturbation would diverge at the singular point. Yet, if this
solution still had some physical relevance, then arbitrary initial
data should be able to excite this mode, and the power spectrum of the
time evolution should show a peak at the corresponding frequency.
However, this is clearly not the case, as can be seen in
Fig.~\ref{fig:ps_poly_1e16}, where we show the late time power spectra
of the time evolution of $u_3$ (left panel) and $h_0$ (right panel)
for model 5. For the fluid variable $u_3$, there is always one single
peak, which varies for different locations $r$ between the boundaries
$\hat\varpi_c$ and $\hat\varpi_s$. There is not even the slightest
trace of a common peak at the expected value of $\alpha = 0.52932$.

For the metric variable $h_0$ (right panel), we essentially observe
the same picture. Here, too, no common peak can be found at the
expected mode frequency $\alpha$, but curiously there is nevertheless
an additional common peak for all locations with its frequency given
exactly by $\hat\varpi_s$. However, this peak does not show up in the
power spectrum of $u_3$, except, of course, directly at the surface.

\begin{figure}
\begin{center}
\vspace*{1cm}
\leavevmode
\epsfxsize=8cm
\epsfbox{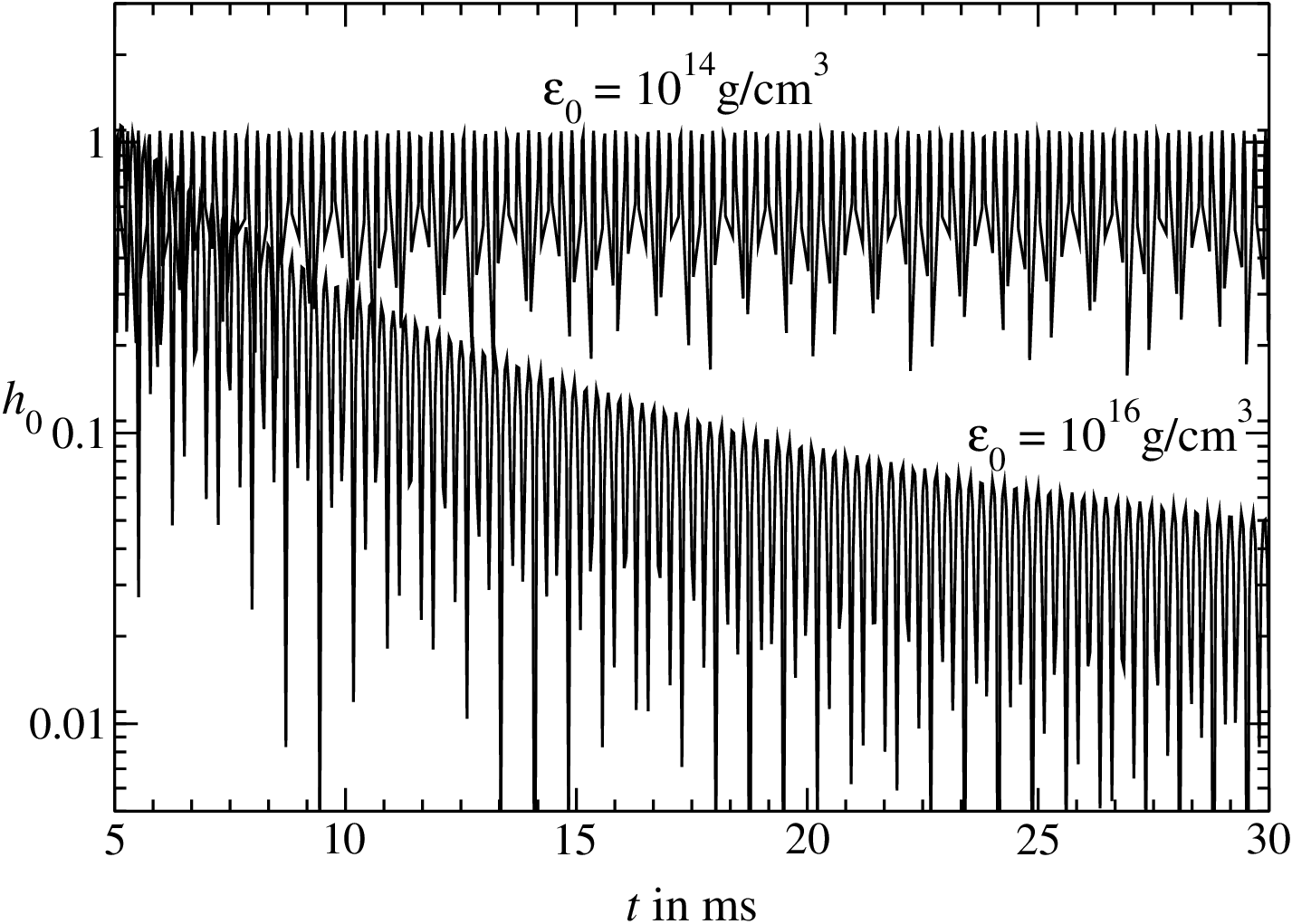}
\vspace*{5mm}
\caption{\label{fig:time_evol}
  Logarithmic plot of the time evolution of $h_0$ outside the star for
  the polytropic models 1 and 5 of Table 1. For model 1, which admits
  a physical mode solution, the oscillation amplitude remains
  constant, and $h_0$ oscillates with the $r$-mode frequency $\alpha$.
  Model 5 does not admit a physical mode solution and here, the
  amplitude of $h_0$ decreases with the decay fitting very well a
  power law with an exponent of -2. The oscillation frequency is given
  by the values of $\hat\varpi_s$.}
\end{center}
\end{figure}

It is obvious that it cannot be a mode solution, since first of all
the eigenvalue code does not give a solution for this particular
frequency $\hat\varpi_s$ or even in the close vicinity. Moreover, the
time evolution shows a quite different behaviour compared to the case
where a physical $r$-mode exists. In Fig.~\ref{fig:time_evol}, we plot
the time evolutions of $h_0$ outside the star for models 1 and 5 of
Table \ref{models}. For model 1, where we have a physical $r$-mode,
after some initial time the amplitude remains constant, whereas for
model 5 the amplitude keeps decreasing with time and in this case fits
very well a power law with an exponent of -2. For model 1, the
dominant oscillation frequency is the corresponding $r$-mode frequency
$\alpha$, whereas for model 5 it is given by $\hat\varpi_s$. In both
cases, the amplitude of the fluid perturbation $u_3$ remains constant
after some initial time. It now becomes clear, why we cannot observe
the common peak at $\hat\varpi_s$ in the fluid spectrum. The spectra
are taken at late times, where the amplitude of $h_0$ and therewith
its influence on $u_3$ has considerably decreased. If we had taken the
spectra at earlier times, we could observe a similar peak in the fluid
spectrum, as well.
  
We have no clear explanation what causes this additional peak, but we
suppose that it comes from the behaviour of the energy density $\eps$
at the surface. The peak is much more pronounced for polytropes with
$n < 1$, since there the energy density $\eps$ has an infinite slope
at the surface. For $n=1$, the slope $\eps'$ is finite and for $n >
1$, it is zero. In the latter case, the peak is strongly suppressed.
Even for uniform density models, one can observe this additional peak,
arising because of the discontinuity of the energy density at the
surface. However, this peak is several orders of magnitudes smaller
than the peak corresponding to the eigenmode, which is always present
for uniform density models, and therefore hard to detect.
  
It is instructive to compare the evolution of the same initial fluid
perturbation for a uniform density model and a polytropic model,
having the same compactness but without the latter admitting a mode
solution. Since in the low-frequency approximation, there is no
radiation which can dissipate the energy of the fluid, the total
energy of the system should be conserved. However, we have observed
that in the polytropic case the amplitude of $h_0$ is constantly
decreasing, hence its initial energy has to be transferred to the
fluid, whereas in the uniform density model, the energy should be
shared between $u_3$ and $h_0$. This is indeed, what can be observed.
In the uniform density case, the fluid amplitude does not change too
much, but in the polytropic case, it shows a quite strong initial
growth, accompanied by the strong decrease of $h_0$.
  
\subsection{Realistic Equations of State}

Having found that for a quite large range of polytropic stellar
models, there do not exist any physical $r$-modes, an obvious question
is, whether or not realistic equations of state do admit physical mode
solutions. To give an answer, we have investigated the collection of
realistic equations of state which have been studied by Kokkotas \&
Ruoff (2001) for the radial modes. The relevant notations, references
and data of the stellar models can be found in there.
  
The results are quite unexpected and seemingly contradictory. When
trying to compute the modes through Eq.~(\ref{eveqn_a}), we find that
for all the equations of state, the frequencies always lie inside the
continuous band. Based on our above discussion, we therefore would
have to discard them as being unphysical.  It then would seem that
none of the existing realistic EOS admits an $r$-mode, at least in the
physically relevant range from about one solar mass up to the
stability limit of each EOS. The surprise is now that the time
evolution does show a quite different picture.  Only for the EOS B
(Pandharipande, 1971), G (Canuto \& Chitre, 1974) and MPA (Wu
et al., 1991) do the evolutions meet our expectations and show the
continuous decrease in the amplitude of the metric variable $h_0$, in
accordance with the polytropic cases without $r$-modes. However, for
all other EOS, the amplitude remains constant after a while,
indicating that there is indeed a mode present. Only when approaching
their respective stability limit do some EOS show a decay of the
amplitude of $h_0$. When obtaining the frequency through Fourier
transformation, we find that it always lies {\em inside} the
continuous spectrum, however, it does {\em not} coincide with the
frequency found from the mode calculation.  Instead, the frequency is
in all cases given by the value of $\hat\varpi$ close to the neutron
drip point.
  
How can we explain this discrepancy with our previous considerations?
First, we would like to stress that it is not a numerical artefact of
the time evolution, for convergence tests corroborate the presence of
this mode. When examining the different EOS, we find that the EOS B, G
and MPA are the softest ones, with a maximal polytropic index in the
high density regime around $n=0.8$.  All others have indices less than
0.8, going down to $n \approx 0.5$ for the EOS I (Cohen et al.~1970)
and L (Pandharipande et al.~1976). From Fig.~\ref{fig:index}, it
becomes clear that it is just for polytropic models with $n \ge 0.8$
that the eigenvalue migrates inside the continuous band, and the
$r$-mode therefore ceases to exist. For models with $n < 0.8$, we
usually can find a physical $r$-mode, but this depends on the
compactness of the model under consideration.
 
\begin{figure}
\begin{center}
\vspace*{1cm}
\leavevmode
\epsfxsize=8cm
\epsfbox{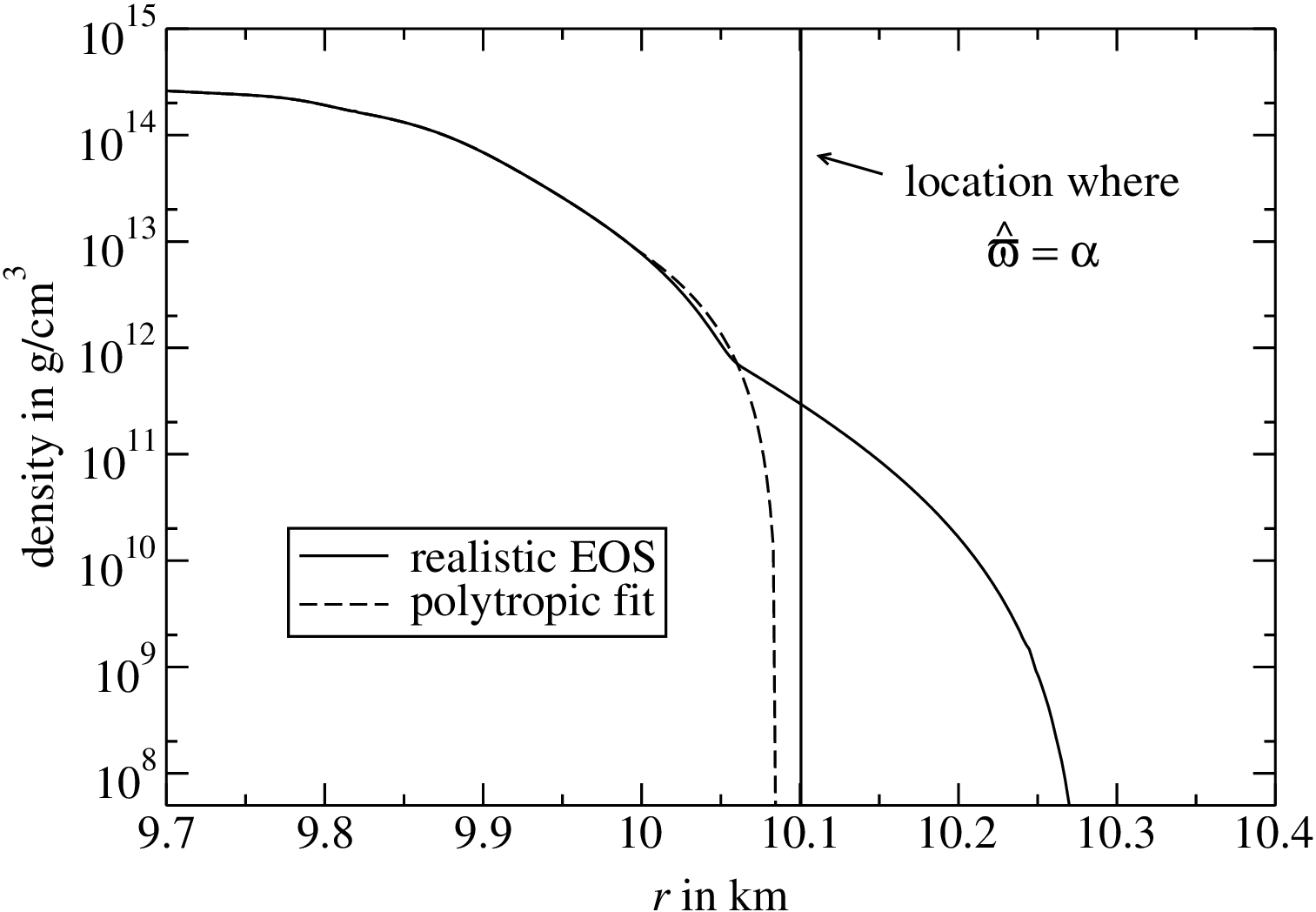}
\vspace*{5mm}
\caption{\label{fig:real_poly}It is shown the density profiles near the
  surface for neutron star model obtained from EOS WFF with a central
  density of $2\times10^{15}$g/cm$^3$. In the one case we include the
  low density part, whereas in the other case this part is replaced by
  a polytropic fit. Also shown is the location where $\hat\varpi =
  \alpha$. For the polytropic fit, it lies outside the star, therefore
  representing a physical mode, but for the complete realistic model,
  it would lie inside.}
\end{center}
\end{figure}

At a density of $10^{14}$g/cm$^3$, the effective polytropic index of
any EOS is given by $n \approx 2$. As the EOS is approaching the
neutron drip point at a density of $\eps \approx 4\times
10^{11}$g/cm$^3$, the EOS becomes softer and softer, i.e. $n$
increases even further. Only for densities below the neutron drip
point does the EOS stiffen again. This structure of the EOS is
responsible of putting a low density layer (the crust) around the high
density core. However, because of its low density compared to the
core, this layer practically does not contribute to the total mass,
its only effect is to slightly increase the radius of the neutron
star. Thus, whether or not the EOS admits a mode, should be determined
solely by the core. To assess this proposition, we can do the
following. We replace the whole part of the EOS below
$10^{14}$g/cm$^3$ by a smooth polytropic EOS with polytropic index of
$n = 2$. By doing so, we obtain a model with practically the same
mass, but a somewhat smaller radius. Computing the modes of the thus
modified model, we find that the $r$-mode frequencies $\alpha$
actually do lie outside the continuous spectrum, if the average
polytropic index of the core is less than 0.8, and the model is not
too compact. However, $\alpha$ lies extremely close to the value of
$\hat\varpi_s$. For softer EOS with an average index of $n \approx
0.8$, we still would not be able to find any physical modes. If we now
go back and restore the outer layer, the $r$-mode frequency should not
significantly change, because of the negligible gravitational
influence of this outer layer. The only effect is the slight increase
of the stellar radius $R$. But with $R$ becoming larger, the value of
$\hat\varpi_s$ also increases and actually becomes larger than the
$r$-mode frequency $\alpha$, which then lies inside the continuous
band. This is shown in Fig.~\ref{fig:real_poly}, where we plot the two
density profiles for a stellar model based on the EOS WFF (Wiringa,
Ficks \& Fabrocini, 1988) and the same model with the low-density
regime replaced by a $n=2$ polytropic fit. For the polytropic fit, the
zero of $\hat\varpi - \alpha$ lies right outside the star, whereas for
the complete realistic model, it is inside the star. In the latter
case, however, the mode still exists, but it cannot have a purely
harmonic time dependence any more. If this were the case, it had to be
a physical solution of Eq.~(\ref{eveqn_a}), but it is clearly not
since the frequency lies inside the continuous spectrum.  We therefore
conclude that this stable oscillation that can be seen in the time
evolution is always a mixture of a mode and the continuous spectrum.
We should mention that in our treatment of realistic EOS, we have
assumed that the neutron stars consists entirely of a perfect fluid,
even in the outer layer. This is certainly not true, instead a neutron
star should have a solid crust, which clearly will modify the above
results. However, this is beyond the scope of this work.

\section{Summary}

We have performed both mode calculations and time evolutions of the
pure axial perturbation equations for slowly rotating stars in the
low-frequency approximation. Although the time independent equation
(\ref{eveqn_a}) represents a singular eigenvalue problem, admitting a
continuous spectrum, it is nevertheless possible to find discrete mode
solutions, representing the relativistic $r$-modes. If the mode
frequency lies outside the continuous spectrum, the eigenvalue problem
becomes regular, and the associated solution represents a physically
valid $r$-mode solution. If the eigenvalue lies inside the continuous
band, the eigenfunction exhibits an infinite slope at the singular
point, which is due to the presence of a $r\log|r|$ term in the
series expansion. Moreover, the corresponding fluid perturbation $u_3$
diverges at the singular point. Therefore, we conclude that these mode
solutions are unphysical, and the only physically valid mode solutions
are the ones where the associated frequencies $\alpha$ lie outside the
range of the continuous spectrum.

We have performed mode calculations for uniform density models, for
various polytropic models and also for a set of realistic equations of
state. In agreement with the results of Lockitch et al.~(2001), we
find that uniform density models generally admit $r$-modes for any
compactness. For polytropic equations of state, however, the existence
of physical $r$-mode solutions depends strongly on both the polytropic
index $n$ and the compactness $M/R$ of the stellar model.  The general
picture is that the smaller $n$ is, which corresponds to a stiffer
equation of state, the larger is the compactness range where one can
find physical mode solutions. For a given polytropic index $n$, one
usually finds physical mode solutions for models with a small $M/R$
ratio. As the compactness is increased, i.e.~as the models become more
relativistic, the mode frequency $\alpha$ decreases and starts
approaching $\hat\varpi_s$.  Eventually it crosses this point and
migrates inside the range of the continuous spectrum, thus becoming
unphysical, and no $r$-mode exists any more.

When considered as a function of $l$, the $r$-mode frequency $\alpha$
is monotonically decreasing. For a uniform density model, $\alpha$
approaches $\hat\varpi_s$ as $l$ is increased. In polytropic models, this
has the effect that it is even harder to find physical mode solutions
for higher values of $l$, since $\alpha$ will much sooner cross the
border $\hat\varpi_s$ of the continuous spectrum. If $l$ is large enough,
it seems that the eigenvalue equation (\ref{eveqn_a}) does not admit
any mode solution at all, not even a singular one.

We have verified our results by explicitly integrating the time
dependent equations. The time evolutions for the models admitting an
$r$-mode can clearly be distinguished from those without a discrete
mode. In the former case, both the fluid perturbation $u_3$ and the
metric perturbation $h_0$ oscillate with a constant amplitude after
some initial time. In the latter case, the amplitude of $h_0$
constantly decreases. The fluid amplitude, however, still remains at a
constant level. This can be explained by a decoherence effect in the
fluid oscillations, since the frame dragging causes each fluid layer
to oscillate with a different frequency. Thus the initially uniform
fluid profile becomes more and more disturbed because the fluid layers
get out of phase, resulting in a continuously weakening of the
strength of the fluid source term in Eq.~(\ref{eqh0}). When a physical
$r$-mode exists, the system can oscillate in a coherent manner.

When turning to realistic equations of state, the mode calculations
yielded only frequencies lying inside the continuous band, therefore
being apparently unphysical. However, for most EOS the numerical time
evolutions revealed the presence of a mode, but with the frequency
still lying inside the continuous band and corresponding approximately
to the value of $\hat\varpi$ at the neutron drip point.  We explained
this seemingly contradictory behaviour by making the core responsible
for the existence of the $r$-mode. For, if we remove the outer layer,
which does not have any significant gravitational contribution, we can
indeed find eigenvalues $\alpha$ which lie outside the continuous
band.  However, they are very close to the upper limit of the
continuous band $\hat\varpi_s$. By adding the additional layer, we
increase $\hat\varpi_s$ such that it now becomes larger than $\alpha$,
which remains basically unaffected. Although now lying inside the
continuous spectrum, the mode still exists, but it will be always
associated with an excitation of the continuous spectrum.  Most of the
realistic EOS do admit $r$-modes in a certain mass range, but only the
stiffest ones admit modes throughout the whole mass range up to the
stability limit. The less stiff ones have a maximal mass model above
which there are no $r$-modes any more, and for the softest EOS, namely
EOS B, G and MPA, there are no $r$-mode for the whole physically
relevant mass range.


It should be kept in mind that all our results concerning the
$r$-modes are obtained within the low-frequency approximation. It
would be clearly much too early to infer any statements about the
existence or non-existence of the $r$-modes in rapidly rotating
neutron stars.  And if the true EOS of neutron stars is rather stiff,
and therefore would already admit $r$-modes within the low-frequency
approximation, then the whole discussion about the singular structure
would be irrelevant. But as the true EOS is not known yet, we cannot
exclude it to be rather soft, and the appearance of the singular
points has to be taken much more seriously. Still, it could still be
seen as a mere artefact of the low-frequency approximation.  But the
work of Kojima \& Hosonuma (1999) indicates that the inclusion of
second-order terms in $\Omega$ even increases the range of the
continuous spectrum, which is responsible for the disappearance of the
$r$-mode. They worked only in the Cowling approximation, but whether
or not the inclusion of more higher order terms and the complete
radiation reaction can restore the existence of the $r$-modes in all
cases is still an open question and deserves further investigation. As
a next step in this direction, we will investigate in a subsequent
paper the full set of axial equations (Eqs.~(\ref{V4}) -- (\ref{u3})),
which contains the radiation reaction.

\section*{acknowledgments}
We thank Nils Andersson, Horst Beyer, John Friedman, Luciano Rezzolla,
Adamantios Stavridis, Nikolaos Stergioulas and Shin Yoshida for
helpful discussions.  J.R. is supported by the Marie Curie Fellowship
No.  HPMF-CT-1999-00364. This work has been supported by the EU
Programme 'Improving the Human Research Potential and the
Socio-Economic Knowledge Base' (Research Training Network Contract
HPRN-CT-2000-00137).

\label{lastpage}

\newpage

\end{document}